\documentclass[12pt]{article}
\usepackage{amsmath}
\usepackage{amssymb}
\usepackage{graphicx}
\usepackage[dvips]{color}
\usepackage{colordvi}

\newlength{\dinwidth}
\newlength{\dinmargin}
\newcommand\ben{\begin{equation}}
\newcommand\een{\end{equation}}
\newcommand\bea{\begin{eqnarray}}
\newcommand\eea{\end{eqnarray}}

\newcommand\nn{\nonumber \\}
\newcommand\half{\ensuremath{frac{1}{2}}}
\newcommand\del{\partial}

\newcommand\bcP{\bar{\cal P}}

\begin{document}
\thispagestyle{empty}
\addtocounter{page}{-1}
\vskip-0.35cm
\begin{flushright}
UK/08-11 \\TIFR/TH/08-47\\
%{\tt hep-th/0203164}
\end{flushright}
\vspace*{0.2cm}
\centerline{\Large \bf Microstate Dependence of}
\vspace*{0.2cm}
\centerline{\Large \bf Scattering from the D1-D5 System}
\vspace*{1.0cm} 
\centerline{\bf Sumit R. Das}
\vspace*{0.7cm}
\centerline{\it Department of Physics and Astronomy,}
\vspace*{0.2cm}
\centerline{\it University of Kentucky, Lexington, KY 40506 \rm USA} 
\vspace*{1.0cm} 
\centerline{\bf Gautam Mandal}
\vspace*{0.7cm}
\centerline{\it Department of Theoretical Physics,}
\vspace*{0.2cm}
\centerline{\it Tata Institute of Fundamental Research,} 
\vspace*{0.2cm}
\centerline{\it Mumbai 400 005, \rm INDIA}

\vspace*{0.5cm}
\centerline{\tt das@pa.uky.edu, mandal@theory.tifr.res.in}

\vspace*{0.8cm}
\centerline{\bf Abstract}
\vspace*{0.3cm}
\vspace*{0.5cm} 
We investigate the question of distinguishing between
different microstates of the D1-D5 system with charges $Q_1$ and
$Q_5$, by scattering with a supergravity mode which is a
minimally coupled scalar in the leading supergravity approximation.
The scattering is studied in the dual CFT description in the orbifold
limit for finite $R$, where $R$ is the radius of the circle on which
the D1 branes are wrapped. Even though the system has discrete energy
levels for finite $R$, an absorption probability proportional to time
is found when the ingoing beam has a finite width $\Delta E$ which is
much larger than the inverse of the time scale $T$. When $R\Delta E
\gg 1$, the absorption crosssection is found to be {\em independent of
the microstate} and identical to the leading semiclassical answer
computed from the naive geometry. For smaller $\Delta E$, the answer
depends on the particular microstate, which we examine for {\em
typical} as well as for {\em atypical} microstates and derive an upper
bound for the leading correction for either a Lorentzian or a Gaussian
energy profile of the incoming beam.  When $1/R \gg \Delta E \gg$ the
average energy gap $\left(1/(R\sqrt{Q_1Q_5}) \right)$, we find that in
a {\em typical state} the bound is {\em proportional to the area of
the stretched horizon,} ${\sqrt{Q_1 Q_5}}$, up to $\log (Q_1Q_5)$
terms. Furthermore, when the central energy in the incoming beam,
$E_0$,  is much smaller than $\Delta E$, the proportionality constant is
a pure number independent of all energy scales.  Numerical
calculations using Lorentzian profiles show that the {\em actual
value} of the correction is in fact proportional to ${\sqrt{Q_1Q_5}}$
without the logarithmic factor.  We offer some speculations about how
this result can be consistent with a resolution of the naive geometry
by higher derivative corrections to supergravity.

\baselineskip=18pt

%%%%%%%%   Gautam's defs  %%%%%%%%%%%%%%

\def\gap#1{\vspace{#1 ex}}
\def\be{\begin{equation}}
\def\ee{\end{equation}}
\def\bal{\begin{array}{l}}
\def\ba#1{\begin{array}{#1}}  %% e.g. \ba{cc}
\def\ea{\end{array}}
\def\bea{\begin{eqnarray}}
\def\eea{\end{eqnarray}}
\def\beas{\begin{eqnarray*}}
\def\eeas{\end{eqnarray*}}
\def\eq#1{(\ref{#1})}
\def\fig#1{Fig \ref{#1}} 
\def\re#1{{\bf #1}}
\def\bull{$\bullet$}
\def\ub{\underbar}
\def\nl{\hfill\break}
\def\ni{\noindent}
\def\bibi{\bibitem}
\def\ket{\rangle}
\def\bra{\langle}
\def\vev#1{\langle #1 \rangle} 
\def\lsim{\stackrel{<}{\sim}}
\def\gsim{\stackrel{>}{\sim}}
\def\mattwo#1#2#3#4{\left(
\begin{array}{cc}#1&#2\\#3&#4\end{array}\right)} 
\def\tgen#1{T^{#1}}
\def\half{\frac12}
\def\floor#1{{\lfloor #1 \rfloor}}
\def\ceil#1{{\lceil #1 \rceil}}

\def\mysec#1{\gap1\ni{\bf #1}\gap1}

\def\bit{\begin{item}}
\def\eit{\end{item}}
\def\benu{\begin{enumerate}}
\def\eenu{\end{enumerate}}

\newpage
\tableofcontents

\section{Introduction and Summary}

The two charge system, particularly in the duality frame in which it
is a bound state of $Q_1$ D1 branes and $Q_5$ D5 branes, has served as
a useful theoretical laboratory for understanding the physics of black
holes in string theory. In fact, this system (in the duality frame in
which this is a fundamental heterotic string with some momentum), 
provided the first evidence that string theory states can
account for black hole entropy \cite{Sen:1995in}.

When the D5 branes are wrapped on a $T^5$ and the D1 branes are
wrapped on an $S^1$ (of radius $R$) contained in the $T^5$, the
microscopic description of the system at low energies is that of a
$(4,4)$ superconformal field theory with a target space which is a
resolution of the orbifold $(T^4)^{N}/S(N),\
N=Q_1Q_5$\cite{Strominger:1996sh, Callan:1996dv, de
Boer:1998ip,David:2002wn,balasub2-ref}.  
In the orbifold limit, the SCFT consists
of $N=Q_1Q_5$ copies of free bosonic fields $X^a$ and their fermionic
partners $\psi^\alpha$, $a=1,..,4; \alpha=1,...,4$, in various twist
sectors corresponding to elements of $S(N)$. Any given element of
$S(N)$ can be described by multiple copies of the cyclic permutation
$Z_n$ \footnote{The $Z_n$ twist acts on the bosonic or fermionic fields so
that $n$ copies are strung together into a ``long string'' which lives
in a circle of radius $nR$.}, for various values of $n=1,...N$ (up to
equivalence). A given twist sector, therefore, corresponds to
specific multiplicities $N_{n,\mu}$ ($N^\prime_{n,\mu}$) of the
$Z_n$ twists acting on the bosons (fermions),
 where $\mu = 1 \cdots 4$ denotes a polarization
index.  The numbers $\{N_{n,\mu}, N^\prime_{n,\mu} \}$ are constrained
to satisfy 
\ben \sum_{n=1}^{N} n\,N_n = Q_1 Q_5 \equiv N, \quad N_n
\equiv \sum_\mu N_{n,\mu} + \sum_\mu N^\prime_{n,\mu}
\label{one}
\een
For periodic boundary condition around the circle, the SCFT
is described by the Ramond sector. The 2-charge system
consists of the various degenerate Ramond ground states, one from 
each twist sector. Thus the entropy of the 2-charge system is 
given by
\ben
S = \log \Omega
\een
where $\Omega$ is total number of
twist sectors, or in other words the
number of possible sets of values of $\{ N_{n,a}, N_{n,\alpha} \}$
subject to the condition (\ref{one}). For large $N$, this is given by
\ben 
S = 2\pi \sqrt{2N} 
\een

In two derivative supergravity, the 
standard description of this system is in terms of a 
string frame metric 
\ben
ds^2 = (f_1(r) f_5(r))^{-1/2} [ -dt^2 + dy^2] +
(\frac{f_1(r)}{f_5(r)})^{1/2} [dx_1^2 + \cdots dx_4^2] +
(f_1(r) f_5(r))^{1/2} [dr^2 + r^2 d\Omega_3^2]
\label{three}
\een
where the harmonic functions are given by
\ben
f_1(r) = 
1+ \frac{r_1^2}{r^2} = 1 + \frac{16 \pi^4 g_s l_s^6 R Q_1}{V r^2}~~~~~~~~
f_5(r) = 1+ \frac{r_5^2}{r^2} = 1 + \frac{g_s l_s^2\,Q_5}{r^2}
\een
We will call this the ``naive'' geometry.
The dilaton and the gauge fields are
\ben
e^{-2\Phi}  =  \frac{f_5(r)}{f_1(r)} ~~~~~~
A_{01234y}  =  \frac{1}{2} (1 - f_5^{-1}) ~~~~~~
A_{0y}  =   \frac{1}{2} (1 - f_1^{-1})
\een
Here $l_s$ is the string length, $g_s$ is the string coupling, $R$ is
the radius of the $y$ direction and $V$ denotes the coordinate volume
of the $T^4$ along the directions $x^1 \cdots x^4$.  The area of the
horizon, at $r=0$, vanishes because the size of the $y$ direction
vanishes here - so that there is no Bekenstein-Hawking
entropy. However, as was shown in \cite{Sen:1995in}, the area of the
stretched horizon, argued to appear from $\alpha'$
corrections, reproduces the microscopic entropy up to a numerical
factor.

More recently it has been realized that once higher derivative terms
in supergravity are included, the naive geometry of similar 2-charge
systems gets significantly modified: a finite horizon develops and the
Wald entropy of this latter geometry precisely agrees with the
microscopic answer \cite{Dabholkar:2004yr,Castro:2007sd}. While this
has not been shown for the D1-D5 system on $T^4$ which we are
considering, it is reasonable to expect that a finite
horizon should again develop and the Wald entropy would again
agree, at least up to proportionality, with the microscopic 
entropy.

The near-horizon geometry of (\ref{three}) is locally $AdS_3 \times
S^3 \times T^4$ with an identification $y \sim y + 2\pi R$. The AdS
scale $\ell$ is given by
\ben
\ell^4 = (r_1r_5)^2 = \frac{16 \pi^4 g_s^2 l_s^8 N}{V} =
\frac{\kappa^2N}{4\pi^3 V}
\label{four}
\een
where $\kappa$ denotes the ten dimensional gravitational coupling
constant, related to ten dimensional Newton's constant $G_{10}$ by
$\kappa^2 = 8\pi G_{10}$.
By virtue of the standard AdS/CFT correspondence, string theory on this
geometry is dual to the (4,4) SCFT  on the resolved
orbifold.

In a different direction, Mathur and collaborators have found smooth
horizon-free solutions of leading order supergravity corresponding to
CFT states of the system \cite{Lunin:2001dt,otherfuzzball}.
These ``fuzzball'' solutions become identical to the naive geometry at
large $r$, but start deviating from it at values of $r \sim
(Q_1Q_5)^{1/6} -$  which is where a stretched horizon should be
located. The fuzzball program has been extended to other systems for
which the leading order supergravity solution has a finite horizon
area, e.g. the 3 charge system in five dimensions \cite{threecharge}.

Geometries with horizons, e.g. the naive geometry (\ref{three}) with a
singular horizon, or those with a regular horizon
\cite{Dabholkar:2004yr,Castro:2007sd} obtained in higher derivative
supergravity, should be in some sense \cite{otherfuzzball}
coarse-grained descriptions of the fuzzball geometries. It is clearly
important to understand the precise meaning of this averaging
process. This issue has been studied from various viewpoints, both in
the present context of the D1-D5 system
\cite{balasub2,Balasubramanian:2007qv, Rychkov:2005ji} as well as in a
similar context of BPS states in $AdS_5$ \cite{Corley:2001zk,
balasub1, Rey:2005cn, Suryanarayana:2004ig,mandal1}. In particular,
\cite{balasub2} and \cite{Balasubramanian:2007qv} studied correlation
functions in the orbifold limit of CFT and showed that for {\em
typical microstates} and short time scales these are independent of
the details of the microstate and that they agree, under AdS/CFT, with
the naive supergravity (BTZ) answers at short time scales. The
agreement does not work for {\em atypical microstates} or long time
scales.

In this paper we investigate this problem from a different point of
view. We consider in detail the scattering of certain low energy
supergravity probes off the D1-D5 system at a finite radius $R$ and
ask how, if at all, we can figure out the microstate-dependence from
the $R$-dependent absorption crosssection. The expectation stems from
the fact that at finite $R$ the CFT spectrum is discrete\footnote{It
is important to distinguish finite $R$ effects, which is our primary
interest here, from finite $N$ effects (see section
\ref{discussion}).} and the nature of the discreteness carries
information about the specific microstate the system is in.

In carrying out this study, we encounter a subtlety, {\em viz.}  that
a system with discrete energy levels never absorbs a monochromatic
incident wave at a constant rate, i.e. the probability of transition
to an excited state never becomes proportional to time. One way to
obtain a constant rate is to consider a limit in which the final
states are part of a continuum, leading to Fermi's Golden Rule. A
second way is to consider an incoherent incident beam with a specified
central energy $E_0$ and an energy spread $\Delta E$ \cite{schiff}. We
will consider the latter method and use $\Delta E$ as a measure of the
resolution of an experiment to probe the discreteness of the system.

The supergravity field we consider is a traceless component of the ten
dimensional metric with both polarizations along the $T^4$ directions,
with zero angular momentum along the transverse 3-sphere.  From the
point of view of the six dimensional theory this mode behaves as a
minimally coupled massless scalar.  We will compute the absorption
cross-section in the orbifold limit of the CFT. This calculation is
almost identical to the absorption by the D1-D5-P system calculated in
\cite{Callan:1996dv,Dhar:1996vu,Das:1996wn}. In these papers, the
absorption (or equivalently Hawking radiation) was calculated in the
large $R$ approximation so that the final states can be considered to
belong to a continuum. In this limit, the microscopic answer agrees
exactly with the supergravity grey body factors in the relevant regime
\cite{Maldacena:1996ix}. In the low energy limit, the cross-section
equals the area of the horizon, which is a special case of a more
general result \cite{Das:1996we}. Ref \cite{Maldacena:1996ix}, in
fact, shows that the agreement persists even when the momentum P in
the $y$ direction vanishes, i.e. in the two charge D1-D5 system whose
leading order supergravity solution has a vanishing horizon area.
Indeed, it has been explicitly shown in \cite{Emparan:1997iv} that the
cross-section vanishes at low energies linearly in the energy.

It is not completely clear why the above agreement between the
orbifold limit calculations and the supergravity answers is exact. We
have nothing to add to the existing discussion of this issue in the
literature (see e.g.  \cite{nonrenormalization}). In this paper we
will take this agreement as an {\em empirical fact} and compute
effects of finite $R$ staying entirely in the orbifold approximation.

In this paper, we will not use the large $R$ approximation,
which requires a more careful calculation of the microscopic absorption
cross-section. There are several length scales in the problem: $R,
\ell, E_0$ and $\Delta E$. The semiclassical calculation of the
cross-section is in the regime where the energy of the incident wave
is much smaller than $1/\ell$. Since we are interested in calculating
the leading corrections to this answer, we will require 
\ben 
E_0, \Delta E \ll \ell^{-1}
\label{energy1}
\een 
Because of orbifolding, the energy gap in a sector 
characterized by a twist $n$ (see \eq{one}) is 
$1/(nR)$. Clearly, when $R\Delta E \gg 1$, the
discreteness of the spectrum is completely invisible and one would
expect an answer completely identical to the semiclassical
cross-section in the naive geometry. This is explicitly demonstrated
in sections (\ref{fourone}) and (\ref{fourtwo}).  Similarly, an
incident wave with a $\Delta E$ much smaller than the smallest 
energy gap in a
given twist sector basically behaves as a monochromatic wave and there
is no time independent absorption rate, as pointed out
earlier. We therefore require
$\Delta E$ to be much larger than the {\em average} energy gap. The latter
may be estimated by noting that in a {\em typical} state, $N_n$ is
approximately given by the  "thermal" distribution for 
large values of $N=Q_1Q_5$ 
\cite{balasub2} \footnote{Strictly speaking, since the LHS
of \eq{five} is an integer, the RHS should be replaced by, e.g.,
the nearest integer. The numerical calculations in Section
\ref{micro} are performed after making such a replacement (the
results do not change appreciably, though). In the following we
will understand \eq{five} with this qualifier.}
\ben 
N_n |_{typical} = \frac{8}{\sinh \beta n}
\label{five}
\een
where $\beta$ is determined by the condition (\ref{one}). For large
$N$, the sum in (\ref{one}) may be replaced by an integral and $\beta$
is approximately given by 
\ben
\beta  \approx \pi \sqrt{\frac{2}{N}}
\label{six}
\een
This leads to an average value of $n$, $\langle n \rangle$, given by
\ben
\langle n \rangle \sim \sqrt{N}
\label{seven}
\een
Thus, for such typical states, the average energy gap is given by 
\ben
\delta E \sim \frac{1}{R\sqrt{N}}
\label{eight}
\een
We will therefore work in the regime
\ben
\frac{1}{R} \gg \Delta
E \gg \frac{1}{R\sqrt{N}}
\label{regime}
\een
We will also require 
\ben
E_0 R \ll  1 
\label{energy2}
\een
It is easy to check that these various regimes are mutually
consistent. As we will show in section (\ref{fourfour}), the
corrections due to discreteness we calculate are suppressed by powers
of $1/(R\Delta E {\sqrt{N}})$.

To lowest order in the gravitational coupling constant, the basic
process of absorption is the creation of a pair of open string modes
moving in opposite directions along the long string of length
$2\pi nR$. For a Lorentzian profile of the incident beam, with 
central energy $E_0$ and an energy width $\Delta E$, we show that the
rate of absorption becomes independent of time for time scales much
larger than the inverse of the energy resolution, $T \gg 1/(\Delta
E)$. This is the regime of validity of Fermi's Golden Rule.  In
addition, the effects of recombination of the open string modes can be
ignored if the time $T$ is {\em smaller} than the time taken by the
pair to meet physically, which is $\pi n R$. For typical states, $n
\sim {\sqrt{N}}$; hence the lower and upper bounds on the time are
consistent with the regime (\ref{regime}) provided
we choose $\frac{1}{R} \gg \Delta E \gg
\frac 1 T \gg \frac{1}{R\sqrt{N}}$. These various time scales are
explained in detailed in section (\ref{fourtwo}).

As shown in \cite{Lunin:2001dt,Giusto:2004ip}, the upper bound on the
time scale is reproduced precisely in the fuzzball picture.  In this
picture, the $AdS$ throat of the naive geometry of (\ref{three}) is
replaced by a capped throat so that all incoming waves eventually get
reflected from the cap.  For a class of microstate geometries
corresponding to a twist sector with all twists equal,
\cite{Lunin:2001dt} and \cite{Giusto:2004ip} showed that the
reflection coefficient whose modulus is unity can be written as a sum
of terms which can be thought to arise from the wave entering the
capped throat region with some probability and suffering multiple
reflections between the mouth of the throat and the cap (the argument
is briefly summarized in Section \ref{lunin-mathur}).  While the
reflection from the cap is perfect, that at the throat is not and
every time part of the wave escapes to the asymptotic region. One
therefore obtains an infinite series of waves, separated by a certain
time delay, which all go back to asymptotic infinity. The time delay
is in fact twice the time taken by the wave to go from the mouth to
the cap, which was computed in \cite{Lunin:2001dt} to be precisely
$\pi n R$. It is clear from this discussion that a particular
microstate will appear to absorb the incoming wave for times which are
smaller than this delay time. Note that this discussion is modified
significantly when we consider a `typical state' where there are long
strings of various lengths. In this case the time delay would be much
larger because of interference effects of waves which get reflected from
long strings of different lengths and therefore have different time
delays \cite{Mathur:2005zp}.

When the observation time lies in the regime discussed above,
the absorption rate is constant and an absorption cross-section can be
defined. In this work, we examine the behavior of this absorption
cross-section as we vary the energy resolution.
We find that for
$R\Delta E \gg 1$ the absorption cross-section is {\em independent of
the specific microstate}, 
\ben
\sigma_{\rho} (E_0, \Delta E) |_{R\Delta E \gg 1} 
= \frac{\kappa^2\,N}{4V} K_\rho(E_0,\Delta E)~\int_0^\infty dE~E~
\rho_{E_0, \Delta E}(E)
\een
where $\rho_{E_0,\Delta E}(E)$ is the energy profile of the incoming
incoherent beam and $K_\rho(E_0, \Delta E)$ is the normalization,
\ben
[K_\rho (E_0, \Delta E)]^{-1} = ~\int_0^\infty dE~\rho_{E_0, \Delta E}(E)
\label{defkrho}
\een
This is in exact agreement with the semiclassical cross-section in the
naive geometry (\ref{three}).

We will be interested in two examples of the energy profiles
\bea
{\rm Lorentzian}:~~~~~~~~~~
\rho_{L;E_0,\Delta E}(E) & = & \frac{E}{[(E-E_0)^2
    + (\Delta E)^2]^2} \nn
{\rm Gaussian}:~~~~~~~~~~~~\rho_{G ;E_0, \Delta E}(E) & = & 
E~{\rm exp} \left[ -\frac{(E-E_0)^2}{(\Delta E)^2} \right]
\label{defprofiles}
\eea
For both these profiles,
$\sigma_{\rho} (E_0, \Delta E) |_{R\Delta E \gg 1} \sim \ell^4\,E_0$ 
for $\Delta E \ll E_0$, 
while for $\Delta E \gg E_0$, 
$\sigma_{\rho} (E_0, \Delta E) |_{R\Delta E \gg 1} \sim \ell^4\,(\Delta
E)$,
The naive geometry is thus detected regardless of the microstate,
{\em without the need for any averaging.}

We then turn to the regime \eq{regime} and calculate the corrections
to the above result by a combination of analytical and numerical
techniques. These corrections arise from the difference between
certain sums which appear in the expression for the cross-section and
their integral approximations. These differences may be bounded from
above using McLaurin integral approximation methods.  The result now
depends on the details of the microstate.  For some special
microstates we find that the $N$ dependence of the correction can be
the same as that of the leading classical result, while for some other
atypical states the correction is suppressed by $e^{-N}$.

Motivated by the expectation that geometries with horizons
should appear as some kind of average over microstates, we go on to 
examine these corrections in detail 
when the microstate is {\em typical}, i.e. a
microstate which approximates a thermal ensemble. For large $N$, the
result in such a microstate would be the same as the result obtained
by {\em averaging} over a thermal (equivalently,
microcanonical) ensemble of microstates.  We find that for such {\em
typical microstates}, the leading correction for $E_0 R \ll 1$ is
bounded as follows (see Section \ref{fourfour}). For the Lorentzian
profile defined in (\ref{defprofiles}) we have
\ben
|| \sigma_{L} (E_0, \Delta E) - 
\sigma_{L} (E_0, \Delta E) |_{R\Delta E \ll 1} ||
=
\frac{\kappa^2 {\sqrt{2}}}{ \pi VR}\,\tilde K_L(\frac{E_0}{\Delta E})\,
L_L(\frac{E_0}{\Delta E})\,{\sqrt{N}}\,
\left[ \frac{1}{2} \log (N) - \log \frac{\pi}{{\sqrt{2}}} + \eta
  \right]
\label{bounds}
\een
where $ 0 < \eta < 1$ and the functions ${\tilde K}_L(x)$ and $L_L(x)$ are 
given by
\bea
\tilde K_L(x) & = & 
\left[ \frac{1}{2} + \frac{x}{2}
  \left( \frac{\pi}{2} + \tan^{-1}x  \right)
  \right]^{-1} \nn
L_L(x)& = &
\frac{x^2 + 1}{(1 + (\sqrt{x^2 +1}
 - x)^2)^2}
\label{def-k-l}
\eea
whereas for the Gaussian profile we have
\ben
|| \sigma_{G} (E_0, \Delta E) - 
\sigma_{G} (E_0, \Delta E) |_{R\Delta E \ll 1} ||
=
\frac{8 \kappa^2 {\sqrt{2}}}{\pi VR}\,\tilde K_G (\frac{E_0}{\Delta E})\,
L_G(\frac{E_0}{\Delta E})\,{\sqrt{N}}\,
\left[ \frac{1}{2} \log (N) - \log \frac{\pi}{{\sqrt{2}}} + \eta
  \right]
\label{bounds-gaussian}
\een
where
\bea
\tilde K_G(x) & = & \left[ e^{-x^2} + x {\sqrt{\pi}} (1+ {\rm erf}(x))
  \right]^{-1} \nn
L_G(x)& = &
(x+{\sqrt{4+x^2}})^2~{\rm exp} \left[ - \frac{1}{4}
  ({\sqrt{4+x^2}}-x)^2 \right]
\label{def-k-l-gaussian}
\eea
These are upper bounds.  For the Lorentzian profile, we estimate the
actual value of the correction numerically, and also obtain a
better analytical estimate for $E_0 = 0$. We find that the
correction is actually proportional to ${\sqrt{N}}$ without the $\log
(N)$ factor, i.e. it is proportional to the entropy. While we have not
performed the numerical calculations for the Gaussian profile we
expect a similar answer in that case as well.

Furthermore, for both profiles, the dominant correction in
(\ref{bounds}) in the regime $E_0 \ll \Delta E$ (in powers of
$(E_0/\Delta E)$ ) is {\em independent of all energy scales} and
simply proportional to $\kappa_5^2 {\sqrt{N}}$ where $\kappa_5^2 =
\kappa^2/(VR)$ is the five dimensional gravitational coupling
constant. This quantity has precisely the form of the area of the
stretched horizon and hence also of the horizon in the solution found
by \cite{Dabholkar:2004yr} and \cite{Castro:2007sd} in higher
derivative supergravity (albeit in a different duality frame).

This is an intriguing result. While we do not have a precise
understanding of this correction in the supergravity side, this result
suggests that it may be possible to obtain this correction from a
semiclassical calculation in these geometries. If this is indeed true,
in our scattering experiment averaging over microstates in the dual
CFT corresponds to the geometries corrected by the higher derivative
corrections. As explained earlier, in our system it does not require
any averaging to obtain the naive geometry .  This is in contrast with
3-charge geometries in five dimensions, where the "naive" geometry has
a finite horizon and an averaging over microstates is necessary for
the absorption cross-section to reproduce this geometry.

We leave a detailed study of the gravity interpretation of our
corrections for future work.  It would also be of interest to compare
the corrections for {\em specific microstates} which we obtain, with
more detailed calculation of propagation of finite width wave packets
in the corresponding fuzzball geometries along the lines of
\cite{Lunin:2001dt} and \cite{Giusto:2004ip}. This is also left for
future work.

Our calculation is closely related to the calculation of correlation
functions in \cite{balasub2}. For observation times which are larger
than the lower bound $1/\Delta E$, the amplitude effectively respects
energy conservation and the resulting cross-section is the Fourier
transform of the imaginary part of a suitable correlation function.
In \cite{balasub2} it is shown that for time scales much smaller than
$R{\sqrt{N}}$, this correlation function becomes microstate
independent to leading order and equals the supergravity result in the
zero mass BTZ black hole (which is essentially
the $AdS_3$ geometry with the identification
$y\sim y+2\pi R$). Note that this upper bound for the time
scale is precisely the time delay in a typical microstate geometry of
the type considered in \cite{Lunin:2001dt} and is, therefore, the
average upper bound of the time for applicability of the Golden
Rule. Our work, however, goes much beyond this and leads to a
calculation of the correction to the leading order result.

In Section 2 we summarize known results of classical absorption in the
naive 2-charge geometry for a monochromatic wave and extend the
calculation to arbitrary incoherent energy profiles. In Section 3 we
review some aspects of the calculation of wave propagation in special
microstate geometries as in \cite{Lunin:2001dt}. Section 4 contains
the main results of our paper: the microscopic probability for
absorption, determination of the time scales for applicability of
Fermi's Golden Rule, and the analytical and numerical results for the
microstate dependent absorption cross-section for energy resolutions
discussed above. Section 5 contains a discussion of our
results. Section 6 contains concluding remarks. In Appendix A we give
details of the semiclassical calculation in the naive geometry and
in Appendix B we detail the derivation of the analytic bounds for
corrections to the cross-section due to discreteness. Appendix
C contains some results for Gaussian energy profiles for the
probe. 

\section{\label{sec:class}Classical absorption by ``Naive'' Geometry}

In this section we will calculate the classical s-wave absorption
cross-section of a massless minimally coupled scalar $\phi$ in the
geometry \eq{three}.  An example of such a scalar is the component
$h_{12}$ of the ten dimensional Einstein frame metric, where the
indices refer to the directions $x^1\cdots x^4$ in (\ref{three}).  Our
result is not new and has been obtained earlier in
\cite{Emparan:1997iv}. However, we include this calculation for
comparison with the calculation of \cite{Lunin:2001dt} which will be
summarized in the next section. The relationship between these two
calculations will be important in understanding the results that
follow.

For a monochromatic
wave with frequency $w$ the scalar field is of the form
\be
\phi = S(r) \exp[iwt].
\ee
It is easy to show that $S(r)$ satisfies  the following wave equation:
\be
[(f_1 f_5)^{1/2}w^2 
+ \frac1{r^3 f_5}\del_r(r^3 f_5(f_1 f_5)^{-1/2}\del_r] S(r)=0
\label{s-wave}
\ee
We will follow the
procedure of \cite{Page:1976df,Dhar:1996vu,Das:1996wn,Das:1996we}.
This involves solving the wave equation in the Far and Near regions, 
defined by
\bea
&&
{\rm Far}:  r \gg w\ell^2
\nn 
&&
{\rm Near}:  r \ll  \ell
\label{regions}
\eea
In the above equation, $\ell$ is defined in (\ref{four}).
This is the radius of the curvature of the near-horizon geometry, which
is  $AdS_3 \times S^3 \times T^4$.
To have an overlap between the two regions we will assume that 
\[
w\ell \ll 1
\]
We will match the solutions in the intermediate region:
\be
w\ell^2 \ll r \ll \ell
\label{intermediate}
\ee
This will determine the ratio of the incoming and outgoing modes at
infinity and therefore 
the probability for absorption of a s-wave. The cross-section is
obtained by folding this with the fraction of a plane wave in the
s-wave. 

The details of this calculation are contained in Appendix \ref{classical1}.
The final result for the low energy absorption cross-section
$\sigma_{cl}$ is
\ben
\sigma_{cl} (w) = \pi^3 \ell^4\,w
\label{classcross}
\een

The expression (\ref{classcross}) is for a monochromatic wave. For our
purposes we will need the absorption cross-section for an incident
incoherent wave characterized by a distribution 
$\rho_{w_0, \Delta w}(w)$ of
frequencies, with 
\ben
\int_0^\infty\,dw~ \rho_{w_0, \Delta w}(w) = 1
\een
Here $w_0$ denotes the peak of the distribution and $\Delta w$ the width.
This cross-section is simply given by
\ben
\sigma_{\rho,classical}(w_0,\Delta w) = 
\int_0^\infty dw\,~ \rho_{w_0, \Delta w}(w)\,\sigma_{cl}(w)
\label{class1}
\een
The integrals in (\ref{class1}) may be evaluated explicitly.
Using (\ref{classcross}) we finally get
for the two profiles defined in (\ref{defprofiles})
\bea
\sigma_{L,classical}(w_0,\Delta w) & = & \pi^3~\ell^4~(\Delta
w)~G_L(w_0/\Delta w)
\nn
\sigma_{G,classical}(w_0,\Delta w) & = & \frac{1}{2} \pi^3~\ell^4~(\Delta w)
~G_G (w_0/\Delta w)
\label{lorentzclass1}
\eea
where
\bea
G_L(x) & = & x + \frac{\left( \frac{\pi}{2} + \tan^{-1} x
  \right)}{1+x \left( \frac{\pi}{2} + \tan^{-1} x
  \right)}  \nn
G_G (x) & = & \frac{2x~e^{-x^2}+{\sqrt{\pi}}(2x^2+1)(1+{\rm erf}(x))}
{e^{-x^2}+x{\sqrt{\pi}}(1+{\rm erf}(x))}
\label{defgfunctions}
\eea

Using the properties of the functions $G_L(x), G_G(x)$ we note that
\bea
\sigma_{L,classical}(w_0,\Delta w) & = & \pi^3 \ell^4
\left[ w_0 + \frac{(\Delta w)^2}{w_0} + \cdots \right]
~~~~~~~~~\Delta w \ll w_0
\label{lorentzclass4}
\\
 \sigma_{L,classical}(w_0,\Delta w)& = & \pi^3 \ell^4  
\left[ \frac{\pi}{2}(\Delta w) + (2 - \frac{\pi^2}{4})w_0 + \cdots
 \right]
~~~~~~~~~~~~~\Delta w \gg w_0 
\label{lorentzclass3}
\eea
while
\bea
\sigma_{G,classical}(w_0,\Delta w) & = & \pi^3 \ell^4
\left[ w_0 + \cdots \right]
~~~~~~~~~\Delta w \ll w_0
\label{lorentzclass5}
\\
 \sigma_{G,classical}(w_0,\Delta w)& = & \pi^3 \ell^4  
\left[ \frac{{\sqrt{\pi}}}{2}(\Delta w) + (4-\pi)w_0 + \cdots
 \right]
~~~~~~~~~~~~~\Delta w \gg w_0 
\label{lorentzclass6}
\eea

\section{\label{lunin-mathur}Wave 
Propagation in a Specific Microstate Geometry}

In this section we summarize the results of \cite{Lunin:2001dt} for
propagation of a massless scalar wave in the geometry which
corresponds to a specific class of microstates of the D1-D5
system. The geometry is a rotating D1-D5 system with angular momentum
$J$ \cite{Balasubramanian:2000rt,Maldacena:2000dr} with a 6
dimensional metric
\bea
ds^2 & = & -\frac{1}{h(r,\theta)}(dt^2 - dy^2) +
h(r,\theta)f(r,\theta) 
\left( d\theta^2 +
\frac{dr^2}{r^2+a^2} \right) \nn 
& - & \frac{2a r_1 r_5}{h(r,\theta)f(r,\theta)}\left(
\cos^2 \theta \,dy\,d\psi + \sin^2 \theta\,dt\,d\phi \right) \nn 
& + & h(r,\theta) \left[ \left( r^2 + \frac{a^2 r_1^2r_5^2
    \cos^2\theta}{(h(r,\theta)f(r,\theta))^2} \right)
  \cos^2\theta\,d\psi^2 + \left( r^2 + a^2 - \frac{a^2 r_1^2 r_5^2
  \sin^2 \theta} {(h(r,\theta)f(r,\theta))^2} 
\right) \sin^2 \theta\,d\phi^2 \right] \nonumber
\eea
where
\ben
f(r,\theta) = r^2 + a^2\cos^2\theta~~~~~~~h(r,\theta) =
\left[(1+\frac{r_1^2}{f(r,\theta)})(1+\frac{r_5^2}{f(r,\theta)})
  \right] 
\nonumber
\een
The radius of the $y$ direction is $R$ and the angular momentum $J$ is
given by 
\ben
J = \frac{1}{2} Q_1Q_5\,\frac{R}{r_1r_5}\,a
\nonumber
\een

For $a=0$ we get back the naive geometry
(\ref{three}) with an infinite throat. For nonzero $a$ the throat is
replaced by a cap.

In \cite{Lunin:2001dt} the wave equation of a massless minimally
coupled scalar was solved in this geometry. Since there is a cap, the
reflection coefficient ${\cal R}$ at infinity satisfies $|{\cal R}| =
1$. However, ${\cal R}$ may be written as an infinite series of terms
which may be interpreted as arising from  the wave  that 
enters the throat and repeatedly undergoes the 
process of reflection by the cap and 
part reflection and part outward transmission at the
throat. For the s-wave and for
\footnote{Note that the
  $\omega$ which appears in \cite{Lunin:2001dt} is equal to $wR$ in
  our notation.} 
\ben
w \ell^2 \ll R~~~~~~~~~~~~~~w^2 (r_1^2+r_5^2) \ll 1
\label{lmathurregime1}
\een
this expansion is 
(see equation (4.24) of \cite{Lunin:2001dt}) 
\ben
R \sim e^{-i\pi\epsilon} - 2\pi^2 \frac{(w\ell)^4}{16}
- 4\pi^2  \frac{(w\ell)^4}{16} \sum_{m=1}^\infty e^{2\pi i m \frac{wRN}{4J} }
\label{reflection}
\een
Here $\epsilon$ us a regulator which is similar to $\nu -1$,
where $\nu$ is as in Section \ref{classical1}.

The expression (\ref{reflection}) 
is an infinite series of terms representing waves with successive
time delays of 
\ben
t_{delay} = 2\pi \frac{\partial}{\partial w}( \frac{wRN}{4J} ) =
\frac{\pi RN}{2J}
\label{timedelay}
\een
In \cite{Lunin:2001dt} this expression was interpreted as follows.
The $m$-th term is the contribution for a wave
which went into the throat, and re-emerged after going back and forth
between the cap and the mouth of the throat $m$ times. The probability
for entering the throat can be then read off from (\ref{reflection})
\ben
{\cal P}_{throat} = 4 \pi^2 \frac{(w\ell)^4}{16}
\een
This is in precisely the same as the probability for {\em absorption}
in the naive geometry, as may be seen by substituting $\mu = 1$ in
(\ref{prob2}).

While the above calculation has been performed for some special
microstates, the lesson is quite general. Since microstate geometries
do not have any horizon, there is no net absorption. However, for
observation times $t_{obs} \ll t_{delay}$ it appears that the system
is absorbing. For large enough $R$ the effective absorption
probability is equal to that by the naive geometry.

The time delay $t_{delay}$ has an important interpretation in the
microscopic model of the D1-D5 system in the orbifold limit.
In this limit, the corresponding microstate is described, in
the notation used in (\ref{one}) by
\ben
N_n = 2J~~~~~{\rm for~}\,n=\frac{N}{2J}
\label{equallength}
\een
and zero otherwise. Therefore in the long string picture this
represents $N/n$ long strings each with winding number $n$ around the
compact circle of radius $R$. Substituting (\ref{equallength}) in
(\ref{timedelay}) we find that
\ben
t_{delay} = \pi R\ n
\label{timedelay2}
\een
which is precisely the time taken by the pair of open strings produced
by the incoming wave to go around the long string and meet each other
again and possibly annihilate to an outgoing mode. This precise
understanding of the time delay is an important ingredient of the
fuzzball picture for such black holes.

\section{\label{micro}Microscopic absorption cross-section}

In this section we perform the microscopic calculation of the
absorption cross-section using a Lorentzian profile for the incoming
wave using a combination of analytic and numerical techniques. The
conclusions based on analytic techniques are valid for a Gaussian
profile: the corresponding results are given in Appendix C.

Our calculation of the microscopic absorption cross-section is
basically a repetition of that in \cite{Callan:1996dv, Dhar:1996vu,
Das:1996wn} for the 3-charge system. However, these papers - as well
as similar calculations for various systems in the literature - deal
with the large $R$ limit. In this limit, the states of the system form
a continuum and Fermi's Golden Rule may be applied in a
straightforward manner. In contrast, we are interested in the effects
of finite $R$ and therefore the discreteness of the states. As is well
known, in the presence of a {\em monochromatic} wave, the transition
probability to some excited state is {\em not proportional to time}
and therefore there is no absorption cross-section. One way to obtain
a constant transition rate is to shine the system with an incoherent
beam with a finite energy width. One of our aims is to determine the
time scales where such a constant rate is obtained. Therefore we need
to be careful about retaining a finite but large time for the
transition process.

In the orbifold limit, the system is equivalent to a collection of
long strings. For a given $N_n$ satisfying (\ref{one}) we have $N_n$
independent long strings with length $2\pi n R$. Let us consider the
interaction of a component of the metric $h_{12}$ where $(1,2)$ denote
two of the directions of the torus in $x^1 \cdots x^4$ direction. From
the point of view of five dimensions, this is a minimally coupled
scalar which we will denote by $\chi (t,y,r,\theta_a,x^i) =
\chi(t,y,z^m,x^i)$, where $z^m,\, m = 1 \cdots 4$ are the four
Cartesian coordinates in the transverse space parametrized by
$(r,\theta_a)$.

As usual we will work in the approximation which ignores brane recoil,
so that the momentum along the transverse direction is not
conserved. The interaction term in the long string action located at
$z^m=x^i=0$ for a winding number $n$ is 
\ben S_{int} =
{\sqrt{2}}\kappa \int_{-T}^T dt \int_0^{2\pi n
R}\chi(t,y,0,0)\,\partial_\alpha X^1\,\partial^\alpha X^2
\label{action}
\een
where $X^1$ and $X^2$ denote the two transverse locations of the long
string and $\alpha = (t,y)$. As argued in \cite{Das:1996wn} the
coefficient in $S_{int}$ is uniquely determined by the equivalence
principle and is independent of the details of our system. This
universality of the interaction coupling is the key to the ability to
derive the numerical factor in the absorption cross-section (or
Hawking radiation rate) for the 3-charge system in \cite{Das:1996wn}.
In general, e.g. for non-minimally coupled modes, 
the coupling is fixed by AdS/CFT correspondence \cite{David:1998ev}.

In (\ref{action}) the field $\chi$ is normalized in the entire nine
dimensional space, while the fields $X^1,X^2$ are normalized in one
spatial direction, $y$. Consider an initial state which is a Ramond sector ground state
of the two dimensional field theory defined by (\ref{action}). 
Consider a bulk field which has zero momenta in the $y$ direction, 
as well as zero angular momentum 
along the transverse $S^3$ composed of 
the angles $\theta_a$. 
To lowest order in the coupling $\kappa$, such a bulk mode with energy
$E$ cannot change the twist sector of the system. The only 
excitation it can produce is a pair of long string modes, one of
which is left moving and the other is right moving. Using standard
Feynman rules the probability for this process is given by 
\ben {\cal
P}_n(E,T) = \frac{4 \kappa^2}{\pi\ V}\frac{1}{ER} \sum_{m=1}^\infty
(\frac{m}{nR})^2 \left[ \frac{\sin
(E-\frac{2m}{nR})T}{(E-\frac{2m}{nR})} \right]^2
\label{transprob}
\een
The momenta of the two $X$-quanta are $\pm (2m)/(nR)$. As expected,
the transition probability does not depend on $R$ when expressed in
terms of the momenta. However, there is 
a factor of $(VR)$ in the denominator because 
the bulk field is normalized in the entire nine dimensional space.
The sum over $m$ is the sum over the momenta of the long string
modes. The expression (\ref{transprob}) denotes the transition
probability due to a long string of winding number $n$. To obtain the
total transition probability for a given state labelled $N_n$ we need
to sum over $n$, 
\ben {\cal P} (E,T) = \sum_{n=1}^N N_n \,{\cal P}_n (E,T) 
\label{transprob2}
\een

\subsection{Infinite $R$ limit\label{fourone}}.

When $R$ is much larger than all other length scales in the problem
(in particular $R \gg 1/E$), the sum over $m$ may be replaced by an
integral over the momenta $p = m/(nR)$. In this limit the expression
(\ref{transprob}) becomes
\ben
{\cal P}_n(E,T) \rightarrow 
 \frac{4 \kappa^2}{\pi\ V}
\frac{n}{E}\int_0^\infty dp\,p^2\,
\left[ \frac{\sin
(E-2p)T}{(E-2p)} \right]^2
\label{transprobcont}
\een
In the limit $T \rightarrow \infty$ the factor
$[\sin (E-2p)T(E-2p)]^2$
has a sharp peak around $E = 2p$ and may be effectively replaced by
$\pi T \delta (E - 2p)$.
The probability then becomes proportional to the time $2T$:
this is the approximation involved in Fermi's Golden Rule.

Performing the integral we get
\ben
{\cal P}_n(E,T) \rightarrow  \frac{4 \kappa^2}{\pi V}
\frac{n \pi T E}{8}
\een
so that the cross-section for absorption for this particular long
string is given by
\ben
\frac{{\cal P}_n(E,T)}{2T} \rightarrow
\frac{\kappa^2\,n\,E}{4V}
\een
The net cross-section is therefore
\ben
\sigma (E) \rightarrow \sum_{n=1}^N N_n \frac{\kappa^2\,n\,E}{4V} = 
\frac{\kappa^2\,E}{4V}\sum_{n=1}^N n\,N_n
= \frac{\kappa^2\,E\,N}{4V}
\label{crosscont}
\een
where we have used (\ref{one}). Note that this answer is independent
of $N_n$ and therefore {\em independent of the particular microstate
  chosen}. This happened because ${\cal P}_n$ became proportional to $n$.
Furthermore this is in exact agreement with the
semiclassical absorption of a monochromatic wave by the naive
geometry,  equation (\ref{classcross}). From the point of view of a
scattering process, the naive geometry is obtained in the extreme long
wavelength limit, regardless of any averaging over microstates.

A more detailed discussion of the validity of the integral
approximation is contained in subsection (\ref{fourfour}).

\subsection{Finite $R$: Time Scales\label{fourtwo}}

For finite $R$, the system has discrete energy levels and the
probability for transition due to a monochromatic wave never becomes
proportional to time. However, for an incoherent incident wave packet
with an energy profile $\gamma (E)$ there will be a regime where the
probability becomes proportional to time and system has a net
absorption \cite{schiff}.  In this subsection we will examine the time
scales when the system displays absorption. The discussion in this
section is closely related to that of applicability of Fermi's Golden
Rule.

The transition probability from an initial state $i$ to a final state
$f$ in the presence of
a monochromatic wave with energy $E$ over a time period $2T$, $\bcP
(E,T)$  is of the form
\ben
\bcP(E,T) = F_{ij}(E)\,\left[ \frac{\sin
(E-\Delta \omega)T}{(E-\Delta \omega)} \right]^2
\een
where $\Delta\omega$ stands for the difference of the energies of the
initial and final states and $F_{ij}(E)$ is a function of $E$
containing matrix elements, phase space factors etc. 
Thus the probability in the presence of an incoherent beam is
\ben
\bcP(T) = \int dE\,\gamma (E)\,
\bcP(E,T) = \int_0^\infty dE\,F_{ij}(E)\gamma(E)\,\left[ \frac{\sin
(E-\Delta \omega)T}{(E-\Delta \omega)} \right]^2
\label{ktwo}
\een
When $T$ becomes large, function $[\sin (E-\Delta \omega)T/(E-\Delta
\omega)]^2$ is sharply peaked at $E=\Delta \omega$, and the width of
the central hump of this function at $E=E_0$ is $1/T$. If the function
$F_{ij}(E)\gamma(E)$ is slowly varying in the region of this hump, we
can replace $F_{ij} (E) \gamma(E) \rightarrow
F_{ij}(\Delta\omega)\gamma(\Delta\omega)$ inside the integral in the
large time limit. This effectively means that under these
circumstances one may make the replacement $[\sin (E-\Delta
\omega)T/(E-\Delta \omega)]^2 \rightarrow \pi T \delta (E - \Delta
\omega)$ 
and
the probability becomes proportional to time $2T$. The criterion of
slow variation of $F_{ij}(E)$ provides a {\em lower bound} for the
time of observation $2T$. 

Consider now the probability of transition of an initial bulk mode
into a state of long string modes when the initial bulk mode has a
Lorentzian energy profile specified by a function $\rho_{L;E_0,\Delta
E}(E)$ (see \eq{defprofiles}). 
The transition amplitude is obtained from
(\ref{transprob}),(\ref{transprob2}) and (\ref{ktwo}),
\ben
{\cal P}_{L}(T) = \frac{4\kappa^2}{\pi VR}\,K_L(E_0,\Delta E)\,
\sum_{n=1}^N N_n \sum_{m=1}^\infty
(\frac{m}{nR})^2 \int_0^\infty dE\,
\frac{1}{[(E-E_0)^2 + (\Delta E)^2]^2}\, \left[ \frac{\sin
(E-\frac{2m}{nR})T}{(E-\frac{2m}{nR})} \right]^2
\label{kone}
\een
In writing down (\ref{kone}) we have interchanged the sum over $m$ and
the integral over $E$. This assumes that at all intermediate steps we
have to work with a cutoff on the upper limit of $m$-summation.
The function $K_L(E_0,\Delta E)$ is the normalization factor 
defined in (\ref{defkrho}).

We want to examine the validity of Fermi's Golden Rule for a given
term in the sum over $n,m$.
For this
we need to determine
the rate of variation of the Lorentzian function which appears in the
integral over $E$. This is easily estimated by calculating the
logarithmic derivative of the function $1/[(E-E_0)^2 + (\Delta
  E)^2]^2$ with respect to $(E-E_0)^2$. Since the integral receives
contributions only when $E_0$ is close to $2m/(nR)$, it is sufficient
to examine this variation at $E=E_0$. This derivative at $E=E_0$ is 
of the order of $1/(\Delta E)^2$. Since the width of the central peak
of $[\sin (E-2m/nR)T/(E-2m/nR)]^2$ is $1/T$, the Golden Rule will hold
when 
\ben
T \gg \frac{1}{\Delta E}
\label{lowerbound}
\een
We have used first order perturbation theory 
in the gravitational coupling $\kappa$. This is justified when
$T$ is below an upper bound set by the inverse of the rate of
transition \cite{baym}. This is usually ensured by requiring the
coupling to be small. In our case, regardless of the coupling one can
obtain a rough upper bound by requiring $T$ to be less than the time
taken for the excitations of the long string to meet and have a chance
to annihilate. Since these excitations move at the speed of light,
this is given by $\pi n R$. Thus we will implicitly assume that
\ben
T \ll nR
\label{upperbound}
\een
As explained in the introduction, this is consistent with the fuzzball
picture.

\subsection{Absorption Cross-section for finite $R$}

When the time $T$ is in the regime determined in the previous
subsection (equations (\ref{lowerbound}) and (\ref{upperbound}) ), 
the transition probability (\ref{kone}) is proportional to
time so that there is an absorption cross-section $\sigma_L
(E_0,\Delta E)$,
\bea
\sigma_L (E_0, \Delta E)& =& \frac{1}{2T}{\cal P}_L (T) = 
\frac{2 \kappa^2}{ VR}\,K_L(E_0,\Delta E)\,
\sum_{n=1}^N N_n  {\tilde{\sigma}}_L(n), 
\nn
{\tilde{\sigma}}_L(n) &\equiv& \sum_{m=1}^\infty
(\frac{m}{nR})^2
\frac{1}{[(\frac{2m}{nR}-E_0)^2 + (\Delta E)^2]^2}\, 
\label{kthree}
\eea
where the function $K_L(E_0,\Delta E)$ is the normalization defined in
(\ref{defkrho}) for the Lorentzian profile
\bea
K_L(E_0,\Delta E)  &&= (\Delta E)^2\, \tilde K_L(E_0/\Delta E)
\nn
\tilde K_L(x) &&=   
\left[ \frac{1}{2} + \frac{x}{2}
  \left( \frac{\pi}{2} + \tan^{-1}x  \right)
  \right]^{-1}
\label{class3-1} 
\eea
Let us first replace the sum over $m$ by an integral. The error
involved in this replacement is evaluated in detail in
Appendix \ref{sec:sum-m} and will be discussed in the following
subsections. Up to an error \eq{error-sigma}, we get
\bea
{\tilde{\sigma}}_L(n)&=& \frac{(nR)^2}{16} \sum_{m=1}^\infty
\frac{m^2}{\left[(m-b)^2 + a^2\right]^2}
\equiv \frac{(nR)^2}{16} \sum_{m=1}^\infty f_L(m)
\rightarrow 
\frac{(nR)^2}{16} 
\int_0^\infty dx\ f_L(x)
\nn
& = & \frac{nR}{8} \left[\frac{E_0}{2(\Delta E)^2} + \frac{E_0^2+(\Delta
    E)^2}{2(\Delta E)^3} \left( \frac{\pi}{2}+
  \tan^{-1}\frac{E_0}{\Delta E} \right) \right]
 \equiv   \sigma_{L,{\rm classical}}(n)
\label{sumoverm}
\eea
Here 
\ben
f_L(x)= \frac{x^2}{((x- b)^2 + a^2)^2},\;\;
b= nRE_0/2, \;\; a=nR\Delta E/2
\label{def-f}
\een
With this, the cross-section \eq{kthree} becomes
\bea
\sigma_L(E_0,\Delta E) & \rightarrow &
\frac{\kappa^2}{ 4V}(\Delta E) G_L(\frac{E_0}{\Delta E})\sum_n n N_n \nn
& = & \frac{\kappa^2}{ 4V} \,N\, (\Delta E) G_L(\frac{E_0}{\Delta E})
\equiv \sigma_{L,{\rm classical}}(E_0,\Delta E)
\label{kfour}
\eea
where the function $G_L(x)$ has been defined in \eq{defgfunctions}. 
In the second line of (\ref{kfour}) we have used (\ref{one}). Clearly
the answer is independent of the specific choice of $N_n$
and therefore  {\em independent of the specific microstate}. Using
(\ref{four}) this is seen to be in perfect agreement with the
supergravity answer in the naive geometry (\ref{lorentzclass1}).

Thus, the dependence of the cross-section on microstates is related to
the difference between the sum over $m$ and its integral
approximation. We now turn to this question in detail.

\subsection{Departure from classical limit for general $E_0$\label{fourfour}}

We saw above (Eqs. \eq{sumoverm}, \eq{kfour}) that the departure from
the classical limit for the cross-section is given by
\bea
\Delta \sigma_L  (E_0, \Delta E)& =&
\sigma_L (E_0, \Delta E) - {\sigma}_{L,\rm classical} (E_0, \Delta E)
\nn
& = & \frac{2 \kappa^2}{ VR}\,K_L(E_0,\Delta E)\,
\sum_{n=1}^N N_n \Delta {\tilde{\sigma}}_L(n)
\label{del-sig}
\eea
where
\be
\Delta {\tilde{\sigma}}_L(n) \equiv {\tilde{\sigma}}_L(n) - {\tilde{\sigma}}_{L,{\rm classical}}(n)
= \frac{(nR)^2}{16} \left( \sum_{m=0}^\infty f_L(m) - \int_0^\infty\ dx\ f_L(x)
\right)
\label{del-sig-n}
\ee
where $f_L(x)$ is as in \eq{def-f}.

Thus, the problem of estimating the correction $\Delta \sigma_L
(E_0, \Delta E)$ reduces to the difference ``(sum $-$ integral)'' 
appearing in \eq{del-sig-n}. This is what we do below, using
the McLaurin integral estimation of sums 
described in  Sec. \ref{integral-approx}.

\subsubsection{Estimation of $\Delta \sigma$}

The McLaurin integral estimation of sums leads
to (see \eq{error-sigma})
\ben
\Delta{\tilde{\sigma}}_L(n) \approx   \eta_3(n)\ \frac1 4 
\frac{E_0^2 + {\Delta E}^2}{({\Delta E}^2 + (\sqrt{E_0^2 + {\Delta E}^2}
 - E_0)^2)^2},\; -1<\eta_3(n)<1
\label{del-sigma-n1}
\een
It is significant that the overall factor of $(nR)^2$ in
(\ref{del-sig-n}) cancels with an inverse factor coming the estimate
of the difference of the sum and the integral. The only $n$ dependence
in (\ref{del-sigma-n1}) is thus in $\eta_3 (n)$.

Using \eq{del-sig}, we therefore get
\ben
\Delta\sigma_{L} (E_0, \Delta E)
= \frac{\kappa^2}{2 VR}\, \tilde K_L(\frac{E_0}{\Delta E})\,
L(\frac{E_0}{\Delta E})\,
\sum_{n=1}^N N_n \eta_3(n)
\label{eta-3-n}
\een
where ${\tilde K_L(x)}$ and $L_L(x)$ have been defined in (\ref{def-k-l}).

Since $|\sum_{n=1}^N N_n \eta_3(n) | < \sum_{n=1}^N N_n$, we have
the following bound 
\bea
\Delta\sigma_{L} (E_0, \Delta E) &<& \Delta\sigma_{L,
{\rm max}} (E_0, \Delta E)
\nn
\Delta\sigma_{L,{\rm max}} (E_0, \Delta E) 
& = & 
\frac{\kappa^2}{2 VR}\, \tilde K_L(\frac{E_0}{\Delta E})\,
L_L(\frac{E_0}{\Delta E})\,
\sum_{n=1}^N N_n
\label{sigma-bound}
\eea
The microstate dependence of this bound is entirely in the 
sum $\sum_{n=1}^N N_n$. 

When $R\Delta E \gg 1$ the integral approximation used in
    (\ref{sumoverm}) is good. This may be seen by examining the
    behavior of the bound (\ref{sigma-bound}). For a given $E_0 R$,
    and large $R\Delta E$ we need the behavior of the functions
${\tilde K}_L(x)\,L_L(x)$ and $G_L(x)$ (defined in (\ref{defgfunctions})) for
    small values of the argument. These are given by
\bea
{\tilde K_L}(x)\,L_L(x) & = & \frac{1}{2} + (1-\frac{\pi}{4})x + O(x^2) \nn
G_L(x) & = & \frac{\pi}{2} + (2- \frac{\pi^2}{4})x + O(x^2)
\label{smallxx}
\eea
Thus 
\bea
\Delta\sigma_{L,{\rm max}} (E_0, \Delta E) & \sim &
    \frac{\kappa^2}{2VR} \frac{\pi}{2} \sum_{n=1}^N N_n \nn
\sigma_{L,{\rm classical}} (E_0, \Delta E) & \sim & 
\frac{\kappa^2 N \pi}{8V} \Delta E 
\eea
Therefore for any $N_n$, i.e. any microstate, the correction goes to
zero compared to the classical answer when $R \Delta E \gg 1$.  This
is what it should be: when $R \Delta E \gg 1$ the energy resolution
is much larger than the level spacing for any microstate. As a result,
the incoming wave practically perceives a continuum
spectrum. 

\subsubsection{$\Delta\sigma$ for a typical state}

Consider now a {\em typical} state of the system for which
$N_n$ is approximately given by (\ref{five}). The sum over $n$ 
can now be estimated (using the same McLaurin approximation
once again:
\ben
\sum_{n=1}^N N_n \approx \frac{8}{\beta} [ - \log(\beta/2) + \eta ],
\; 0< \eta < 1
\label{sum-over-n}
\een
For large $N$, $\beta$ is determined in (\ref{six}). We thus
obtain
\ben
\Delta\sigma_{L,{\rm max}} (E_0, \Delta E)
= \frac{\kappa^2}{2 VR}\, \tilde K_L(\frac{E_0}{\Delta E})\,
L_L(\frac{E_0}{\Delta E})\,\frac{2{\sqrt{2}}}{\pi}\,{\sqrt{N}}\,
\left[ \frac{1}{2} \log (N) - \log \frac{\pi}{{\sqrt{2}}} + \eta
  \right]
\label{genestimate}
\een
Note that this is expected to be an upper bound on the correction.
In Section \ref{sec:e0=0} we will provide a more refined estimate for
$E_0=0$ based on an exact evaluation of the sum over $m$ and find that
in this case the $\log(N)$ term is not present. For $E_0 \neq 0$ our
numerical analysis also indicates that the correction is simply
proportional to ${\sqrt{N}}$.

It is worth pointing out that the emergence of ${\sqrt{N}}$ is tied to
the fact that the only $n$ dependence of $\Delta
{\tilde{\sigma}}_L(n)$ is in the number $\eta_3 (n)$ which always lies
between $\pm 1$.

A similar bound will be derived for an incoming Gaussian profile in 
Appendix C.

\subsubsection{Analysis of the bound}\label{boundanalysis}

It is useful to analyse the expression (\ref{genestimate}) for $E_0
\ll \Delta E$ and for $E_0 \gg \Delta E$, and compare the result with
the leading classical answer (\ref{kfour}). 

\mysec{$E_0 \ll \Delta E$}

For $E_0 \ll \Delta E$ we need to use the small-$x$ expansion of the
functions ${\tilde K_L}(x)\,L_L(x)$ and $G_L(x)$ given in (\ref{smallxx})
This leads to the following expansions for large $N$
\bea
\sigma_{L,classical}(E_0, \Delta E) & = & N \frac{\kappa^2}{4VR} (R\Delta E)
\left[ \frac{\pi}{2}  + (2- \frac{\pi^2}{4})\frac{E_0}{\Delta E}
+ O((E_0/\Delta E)^2) \right] \nn
\Delta\sigma_{L,{\rm max}} (E_0, \Delta E) & = & {\sqrt{N}}\, 
\frac{\kappa^2}{2 VR}\,\frac{2{\sqrt{2}}}{\pi}\,
\left[ \frac{1}{2} + (1-\frac{\pi}{4}) \frac{E_0}{\Delta E} +  
O((E_0/\Delta E)^2) \right]\; 
\left[ \frac{1}{2} \log (N) - \log \frac{\pi}{{\sqrt{2}}} + \eta
  \right]
\nn
\label{smallx}
\eea 
It is clear from (\ref{smallx}) that compared to the classical answer,
our estimate for the correction
is suppressed by a factor of $\frac{1}{R{\sqrt{N}}\Delta E}$ term by
term in an expansion in $\frac{E_0}{\Delta E}$. Perhaps more
significantly the leading correction is {\em independent of the energy
scales}. Up to the $\log(N)$ factor, it is given by 
\ben
\Delta\sigma_{L,{\rm max}} (E_0, \Delta E) |_{E_0 \ll \Delta E} 
  = \kappa_5^2\ \frac1{\sqrt 2 \pi} {\sqrt{N}}\; 
\left[ \frac{1}{2} \log (N) - \log \frac{\pi}{{\sqrt{2}}} + \eta
  \right]
\een
where $\kappa_5^2$ is the five dimensional gravitational coupling
\ben
\kappa_5^2 = \frac{\kappa^2}{VR}
\een
Note that the maximum value of
$\Delta\sigma_L$ is consistent with \eq{crosssection-error-typical-1-txt}
for the $E_0=0$ case.

The entropy of the system is proportional to ${\sqrt{N}}$ so
that (up to the $\log\ N$ factor) 
\ben
\Delta\sigma_{L,{\rm max}} (E_0, \Delta E) |_{E_0 \ll \Delta E} 
\propto A_H
\label{correctsmallE0-l}  
\een
where $A_H$ is the area of the stretched horizon.
If results similar to other 2-charge systems also hold in our case,
this is also the
area of the horizon of the higher derivative corrected black hole
geometry \cite{Dabholkar:2004yr,Castro:2007sd}. 
We will soon see that this leading correction is {\em
negative}.  

\mysec{$E_0 \gg \Delta E$}

For $E_0 \gg \Delta E$ we need to use the large-$x$ expansions of the
functions ${\tilde K_L}(x)\,L(x)$ and $G_L(x)$. These are
\bea
{\tilde K_L}(x)\,L_L(x) & = & \frac{2x}{\pi} + \frac{1}{\pi x} + O(1/x^2) \nn
G_L(x) & = & x + \frac{1}{x} + O(1/x^2)
\eea
This leads to the following expansions 
\bea
\sigma_{L,classical}(E_0, \Delta E) & = & N\ \frac{\kappa^2}{4VR}
(RE_0)
\left[ 1 + \left( \frac{\Delta E}{E_0} \right)^2
+ O(\frac{(\Delta E)^3}{E_0^3}) \right] \nn
\Delta\sigma_{L,{\rm max}} (E_0, \Delta E) & = &{\sqrt{N}}\, 
\frac{\kappa^2}{2 VR}\,\frac{2{\sqrt{2}}}{\pi}\,
\frac{2E_0}{\pi \Delta E}
\left[ 1 + \frac{1}{2} \left( \frac{\Delta E}{E_0} \right)^2  
+ O(\frac{(\Delta E)^3}{E_0^3}) \right]\;
\left[ \frac{1}{2} \log (N) - \log \frac{\pi}{{\sqrt{2}}} + \eta
  \right]
\nn
\label{largex}
\eea 
Once again, terms in $\Delta\sigma_{L,{\rm max}} (E_0, \Delta
E)$  
are suppressed by a factor of $\frac{1}{R{\sqrt{N}}\Delta E}$ compared
to the classical answer term by
term. The leading term is proportional to the area of the stretched
horizon (up to the $\log\ N$ factor), 
but with an additional factor of $\frac{E_0}{\Delta E}$
\ben
\Delta\sigma_{L,{\rm max}} (E_0, \Delta E) |_{E_0 \gg \Delta E} 
\propto A_H \frac{E_0}{\Delta E}
\label{correctlargeE0-l}
\een

\subsubsection{Numerical estimate}

In this subsection we perform the summations in \eq{del-sig}
and \eq{del-sig-n} numerically. 

We attach some numerical plots of 
$\Delta \sigma_L(E_0,\Delta E)/\sqrt N$ 
vs $N$ for $A \equiv \pi R\Delta E=.05$ (chosen
within the regime \eq{regime}) and various values of $E_0 R$.

\begin{figure}[ht!]
\centering
\includegraphics[scale=1]{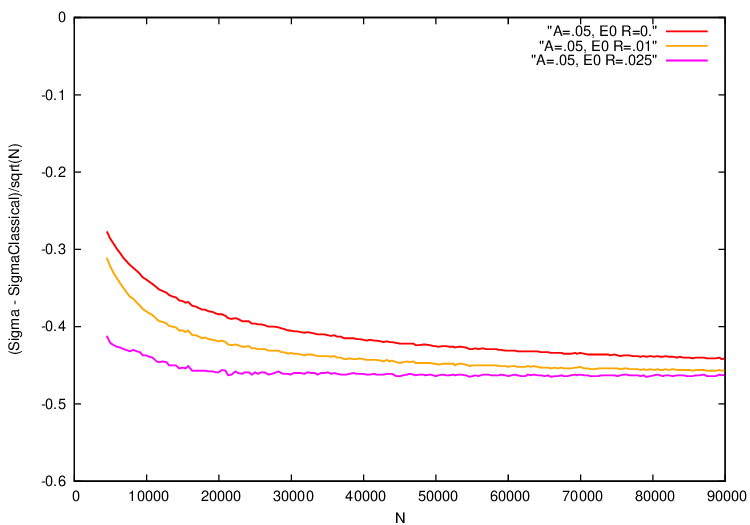}
\caption{
We have plotted $\Delta \sigma_L(E_0,\Delta E)$ (in units of
$\kappa_5^2$) on the
$y$-axis  which is calculated using \eq{del-sig-n} and
\eq{del-sig}. For $N_n$, 
we have used \eq{five} in which the quantity in the RHS
is replaced by its nearest integer. $A= \pi\,\Delta E R$.   
All plots have $E_0R \le 0.25$. 
}
\label{fig:small-e0}
\end{figure}
Figure \ref{fig:small-e0} shows that $\Delta\sigma_L$ is negative for
sufficiently small $E_0$ and that for large $N$ the curves behave as
$\sqrt{N}$.

\begin{figure}[ht!]
\centering
\includegraphics[scale=1]{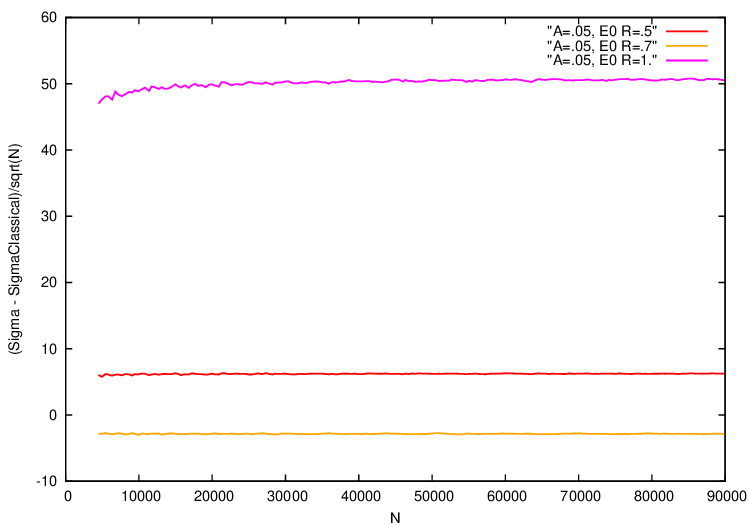}
\caption{In these plots  $E_0 R \ge .5$. Other things are as in Figure
\ref{fig:small-e0}.}
\label{fig:large-e0}
\end{figure}
Figure \ref{fig:large-e0} show that for larger values of $E_0$,
$\Delta\sigma_L$ can be both positive and negative.

Of course, the calculation of the classical cross-section itself is
valid for $E_0 R \ll 1$, together with the other conditions
(\ref{energy1})-(\ref{energy2}) so perhaps only the first set of
plots, $E_0 R\le .25$, are relevant.

\subsubsection{Atypical states}\label{atypical1}

So far we have only considered typical states with $N_n$ given
approximately by the thermal distribution.

Let us now briefly consider some states with atypical $N_n$. 
In particular let us suppose that $N$ can be represented
as a product $N = p\ q$, where $p,q$ are integers. We will
consider
\be
N_n = p \ \delta_{n,q}
\label{same-long}
\ee  
This represents $p$ cycles, each of length $q$.

In this case, \eq{del-sig} simply becomes ($\kappa_5^2 = \kappa^2/(VR)$)
\bea
\Delta \sigma_L  (E_0, \Delta E) &=&
2p\ \kappa_5^2\ K_L(E_0,\Delta E) \Delta {\tilde{\sigma}}_L(q)  
\label{del-sig-atyp}
\eea

Since $\sum_n N_n = p = N/q$ in this case, the bound on the correction
(\ref{sigma-bound}) becomes
\ben
\Delta \sigma_{L,{\rm max}}(E_0, \Delta E) |_{(p,q)} = 
\frac{\kappa^2}{2 VR}\,\frac{N}{q} \tilde K_L(\frac{E_0}{\Delta E})\,
L_L(\frac{E_0}{\Delta E})
\label{atypicalbound}
\een

Let us consider the two extreme cases:

\begin{itemize}

\item Untwisted sector $(p,q)=(N,1)$, i.e.
$N_n= N \delta_{n,1}$ which implies $N$ short cycles,
each of length 1. In this case $\Delta \sigma_{L,{\rm max}}
(E_0, \Delta E) |_{(N,1)}$ in (\ref{atypicalbound}) is proportional to
$N$ just like the classical answer. An analysis similar to that of
Section (\ref{boundanalysis}) then shows that  the finite $R$
corrections are now suppressed by a factor of $1/(R \Delta E)$
compared to the classical result.

%We have
%\[
%\Delta \sigma_L  (E_0, \Delta E) =
%2N\ \kappa_5^2\ K_L(E_0,\Delta E) \Delta \sigma(1)\  \propto\ N
%\]
%In the last step we have used the fact that $\Delta \sigma(1)$,
%as defined by \eq{del-sig-n} and \eq{def-f}, does not contain any 
%factors of $N$.

%{\em Hence the correction to
%the cross-section is as big as the classical limit itself.}

\item Maximally twisted sector $(p,q)=(1,N)$, i.e. $N_n
= \delta_{n,N}$ which implies one long cycle of length $N$:
In this case, $\Delta \sigma_{L,{\rm max}}
(E_0, \Delta E) |_{(1,N)}$ in (\ref{atypicalbound}) is {\em
  independent of $N$} and would be suppressed at least by a power of
$1/(NR\Delta E)$.

%Here
%\be
%\Delta \sigma_L  (E_0, \Delta E) =
%2 \kappa_5^2\ K_L(E_0,\Delta E)\,\Delta \sigma(N) 
%\label{atyp-N}
%\ee
%The $N$-dependence of the last factor is complicated.
%Using \eq{error-sigma} in \eq{atyp-N} we get the following bound 
%\be
%\Delta \sigma_{L,{\rm max}}  (E_0, \Delta E) =
%2 \kappa_5^2\ K_L(E_0,\Delta E)\, 
%\left(
%\eta_3(N) \frac1 4 
%\frac{E_0^2 + {\Delta E}^2}{({\Delta E}^2 + (\sqrt{E_0^2 + {\Delta E}^2}
% - E_0)^2)^2}
%\right)
%\label{atyp-N-est}
%\ee
%
In fact, a numerical estimate shows that $\Delta \sigma_L$
decays exponentially with $N$. To do the numerical computation,
we rewrite \eq{del-sig} for this case as 
\be
\Delta \sigma_L  (E_0, \Delta E) 
= \kappa_5^2\ \frac{N^2}{8}\ (R\ \Delta E)^2\ \tilde
K_L(E_0,\Delta E)\ \left(\sum_{m=0}^\infty\ f_L(m)
- \int_0^\infty\ dx\ f_L(x) \right)
\label{max-twisted-gen-e0}
\ee
The result is
shown for $E_0=.01, A=\pi R\ \Delta E=.05$ in Figure \ref{nsum-3}. 
\begin{figure}[ht!]
\centering
\includegraphics[scale=1]{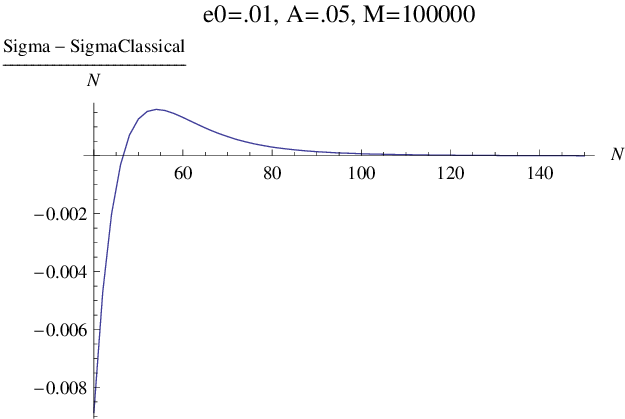}
%\hspace{3ex}
%\includegraphics[scale=.6]{c10b.eps}
\caption{We show $(\sigma-\sigma_{L,{\rm classical}})/N$
vs $N$ in units $\kappa_5^2$. $M$ denotes the upper limit 
in the sum over $m$. $A= \pi R \Delta E,
e_0=R E_0$. The McLaurin upper bound \eq{atypicalbound}
in this case is 0.0278431.}
\label{nsum-3}
\end{figure}
This is consistent with the behaviour \eq{max-twist-atyp-e-0} for
$E_0=0$, {\em viz.}  $\Delta\sigma_L \sim N\,\exp[-N]$.

\end{itemize}

We have performed numerical calculations also for general $p,q$.  From
\eq{atypicalbound}, we find that the correction to the classical
crosssection vanishes for this case also for $R\Delta E \gg
1$. Numerically, we find that the correction vanishes exponentially
with large $q$ and furthermore for values of q
beyond  a certain value depending on $E_0$ and $\Delta E$,
the correction changes sign from negative to positive.

\subsection{\label{sec:e0=0}The cross-section for $E_0=0$}

We now examine the expression (\ref{kthree}) for the case
$E_0 = 0$. Even though the peak of the wave packet is at zero energy,
the spread $\Delta E$ is non-zero, so that the wave packet can sample
several energy levels of the system.

If we were to specialize \eq{sigma-bound} to $E_0=0$, we would
get (using $\tilde K_L(0)=2, L(0)=\frac14$)
\be
\Delta\sigma_{L,{\rm max}} (E_0=0,\Delta E)
= \frac{\kappa^2}{8 VR}\sum_{n=1}^N N_n
\label{crosssection-error-e0-0}
\ee
For a thermal state, this would become 
\be
\Delta\sigma_{L,{\rm max}} (E_0=0, \Delta E)\approx 
- \frac{\kappa^2}{ VR}\,\frac{1}{\beta}
\, \left[\log(\beta/2) - \eta\right]
\label{cross-inferior}
\ee

However, for $E_0=0$ we can get a better estimate, since
 the sum in
(\ref{sumoverm}) can be performed analytically:
\[
\sum_{m=1}^\infty \frac{m^2}{(m^2+a^2)^2} = \frac{\pi}{4a} \left[ 1 + 
H(2\pi a) \right]
\]
where
\be
H(x) \equiv 
\frac{2}{(e^x-1)^2} \{ (1-x)\ e^{x}-1 \}= 
\frac{d}{dx}\,\frac{2x}{e^x-1} 
\label{crosszero2}
\ee
It is straightforward to check that $ -1 \leq H(x) \leq 0$ for 
$0 \leq x \leq \infty$. Therefore the correction is always negative,
though the total sum is of course positive.

Using $a= nR\Delta E/2$ we get
\be
{\tilde{\sigma}}_L(n) = \frac{\pi\ R}{32 \Delta E}\,n\, \left[ 1 + 
H(n\,\pi R\Delta E) \right]
\label{sigma-n-e0-0}
\ee
The ``1'' in the square bracket above gives the classical
expression
\be
{\tilde {\sigma}}_{L,{\rm classical}}(n)=  \frac{\pi\ R}{32 \Delta E}\,n\, 
\label{sigma-class-n-e0-0}
\ee
which can be verified by putting $E_0=0$ in the expression
for ${\tilde{\sigma}}_{L,{\rm classical}}(n)$ in \eq{sumoverm}. We will
see in the next subsection that for typical states the
``1'' term corresponds to the ``continuum  limit'' $R
\gg 1/(\Delta E)$.

Summing over $n$ as in \eq{kfour}, we get
\[
\sigma_{L, {\rm classical}}(0, E_0)=
\frac{\pi \kappa^2}{8V}(\Delta E)
\]
which agrees with the classical answer $\sigma_{\rho,classical}$ 
from appropriate naive geometry, as in equation (\ref{lorentzclass3}).

Using \eq{sigma-n-e0-0} and   \eq{sigma-class-n-e0-0} 
we get 
\be
\Delta{\tilde{\sigma}}_L(n) = {\tilde{\sigma}}_L(n)- 
{\tilde{\sigma}}_{L,{\rm classical}}(n)
= \frac{\pi\,R}{32 \Delta E}\,n\,  
H(n\,\pi R\Delta E)
\label{del-sig-n-e0=0}
\ee
Using this in (\ref{del-sig}) and using $K_L(0,\Delta E) = 2(\Delta E)^2$
we get the following {\em exact expression}, 
\ben
\Delta \sigma_L(0,\Delta E) = \frac{\pi \kappa^2}{8V}(\Delta E) 
\sum_{n=1}^N g(n),\;\; 
g(n)=  n\ N_n\  H(\pi n R \Delta E)
\label{crosszero1}
\een
As pointed out above $\Delta \sigma_L (0, \Delta E) \leq 0$. 

\subsubsection{$\Delta\sigma$ for  a typical state}

The microstate dependence is in the above sum over $n$.
We will now estimate this sum for {\em a typical microstate} at large
$N$. As mentioned before, in this case the occupation numbers $N_n$
has a thermal distribution given by (\ref{five}). Therefore
the function $g(x)$ in \eq{crosszero1} becomes
\be
g(x)= x\ H(A\ x)\ \frac{8}{\sinh(\beta x)}, \; \beta=\pi\sqrt{\frac 2 N},
\; A= \pi R (\Delta E)
\label{g-typical}
\ee
We now again use the method of Sec. \ref{integral-approx} to 
obtain the estimate
\be
\sum_{n=1}^N g(n) =  \int_1^N dx\ g(x) - \eta g(1),\; 0< \eta<1,
\label{mclaurin-increasing-2}
\ee
To find this we have used \eq{mclaurin-increasing} on the function $-g(x)$
which is positive and monotonically increasing for $x>0$. 

\mysec{Case: $R\ \Delta E \gg 1$}

In this case both terms in the RHS behave as $\exp(-\pi R\ \Delta E)$.
Hence 
\be
\Delta\sigma_L \sim \exp(-\pi R\ \Delta E)
\label{A.gt.1}
\ee
and the cross-section becomes classical.

\mysec{Case: $1/\sqrt N \ll  R\ \Delta E \ll 1$}

In order to obtain non-trivial corrections to the classical
cross-section from the naive geometry, we therefore must choose $ R\
\Delta E \ll 1$. Following the considerations leading to \eq{regime}
we will in fact estimate the RHS \eq{mclaurin-increasing-2} in the
regime \eq{regime}.

It is easy to calculate $g(1)$:
\[
g(1) = \frac{1}{\sinh(\beta)}\,\frac{16}{(e^{A}-1)^2}\,\{ (1-A)e^A-1 \}
\approx \frac{16}{\beta}\ (e^{A}-1)^{-2}\,\{ (1-A)e^A-1 \}
\]
For $A \ll 1$, we have 
\be
g(1) \approx  -\frac{8}{\beta}  
\left(1- \frac A 3 + O(A^3)\right)
\label{g-1}
\ee
The integral over $g(x)$  in \eq{mclaurin-increasing-2},
unfortunately, cannot be computed exactly. However,
in the regime of interest here, it can be approximately
evaluated, as follows.

For $A< 1,\  1/A > 1$. Note that the function $H(Ax)$ falls off
as $\sim \exp[-p]$ beyond $x= p/A> 1$. By choosing
$p$ sufficiently large we can ignore $g(x)$ beyond this
value. Hence we can approximate
\[
\int_1^N dx\ g(x) \approx \int_1^{p/A} dx\ g(x)
\]
Also in this range $\beta x \ll  1$, hence 
$x/ \sinh(\beta x) \approx 1$. Using these two ingredients,
we get
\be
\int_1^N dx\ g(x) \approx  \frac{16}{\beta\, A} \,
\left[\frac{p}{e^p-1}- \frac{A}{e^{A}  -1}\right]
\approx -\frac{16}{\beta\, A}
\label{g-int}
\ee
In the last step we have used $A \ll 1$ again. 

Combining \eq{g-int} and \eq{g-1}, and using the
value of $\beta$, we get the final expression
\be
\Delta\sigma_{L} (E_0=0,\Delta E)
\approx 
- \sqrt N\ \kappa_5^2 \  \frac{\sqrt 2}{\pi}\; 
\left[1 - \frac{\eta' \pi R (\Delta E)}{2}
\left(1- \frac{\pi R \Delta E}{3} + O((R\Delta E)^3)\right) \right]
\label{crosssection-error-typical-1-txt}
\ee
where in the last step we have used the fact 
that $\frac{\kappa^2}{VR} = \kappa_5^2$. This is a much more improved
estimate compared to (\ref{smallx}). Since for large $N$ we have
$\beta \sim 1/{\sqrt{N}}$, the leading term in our regime of
interest (\ref{regime}) (which includes $R\Delta E \ll 1$ 
is now seen to be proportional to $A_H$ {\em without any log factor}.

We have verified the above conclusions numerically. A sample
plot is given in Figure \ref{plot:e0-typ}.

\begin{figure}[ht!]
\centering
\includegraphics[scale=1.4]{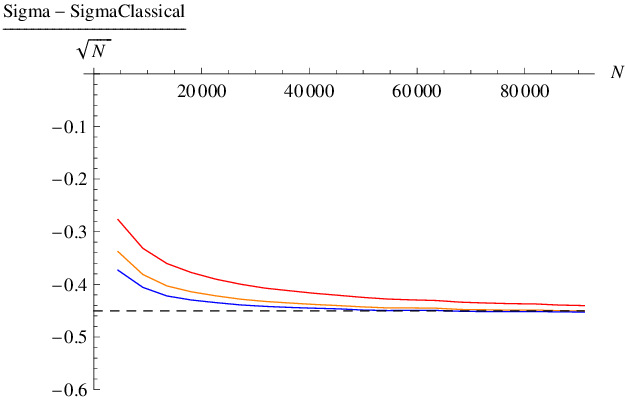}
\caption{$\Delta\sigma_L(0,\Delta E)/\sqrt N$
is numerically computed, using
\eq{crosszero1}, and plotted in units of
$\kappa_5^2$ for various values of
$A= \pi R \Delta E$. The Red, Orange, Blue and Magenta curves  
(which go successively down) 
represent $A= .05, .075, .1$ and .5 respectively.
The curves asymptote to constant values ($-\sqrt 2/\pi\approx -.45$ 
for small $A$ and increases for larger values of $A$) and are
consistent with the analytic estimate 
\eq{crosssection-error-typical-1-txt}.}
\label{plot:e0-typ}
\end{figure}

\subsubsection{Atypical state}\label{atypical2}

We now evaluate such departures for $E_0=0$ in {\em atypical} states:
which we again take as in \eq{same-long}.

The expression \eq{del-sig-atyp} is valid once again, except
that we now have an explicit expression for $\Delta{\tilde{\sigma}}_L(n)$
in terms of the function $H(\pi\ n\ R\ \Delta E)$, see
Eq. \eq{del-sig-n-e0=0}. With this, we get
\be
\Delta \sigma_L  (E_0, \Delta E) =
\frac{\pi}{8}\ \kappa_5^2\ (R \Delta E)\,
N\ H(q\ A) \ \propto  \  N\ H(q\ A)
\label{del-sig-atyp-e=0}
\ee
Here $ A=\pi \ R\ \Delta E$.  

Thus in the maximally twisted case ($(p,q)=(1,N)$) we have
\be
\Delta \sigma_L \ \propto \  N \exp[-N]
\label{max-twist-atyp-e-0}
\ee
whereas in the  untwisted case ($(p,q)=(N,1)$) we have
\[
\Delta \sigma_L \ \propto \  N 
\]
Again, as in the case of general $E_0$, the correction is
as large as the classical limit.

\section{\label{discussion}Discussion of Results}

The type of corrections to the microscopic cross-section we have
calculated in this paper are due to the finite size of the circle on
which the D1 branes are wrapped. Note that this is distinct from
corrections due to finite $N$. In some cases, e.g. typical states,
these corrections are also suppressed by a power of
$1/{\sqrt{N}}$. However in other states, e.g. the untwisted sector,
these corrections are not suppressed by any power of $N$. In all
cases, however, they are suppressed by a power of $1/(R\Delta E)$.

To obtain these corrections one needs to perform two sums: (i) the sum
over discrete values of the momenta and (ii) the sum
over twists (labelled by $m$ and $n$ respectively in \eq{kthree}).

For $R \Delta E \gg 1$ the spectrum of the system is practically
continuous, and the sum over momenta can be replaced by an
integral. Once this is done, the sum over twists is trivial since it
appears as $\sum_n n\,N_n$ which is by definition $N$ see
\eq{kfour}. This is why the resulting cross-section is {\em
independent of the particular microstate the system is in}. The result
is also in exact agreement with the semiclassical cross-section
obtained in the naive geometry.

The bounds on the corrections to this classical result have been
obtained (see Section \ref{fourfour}) by using McLaurin estimates for
the difference between the discrete sums (i) and (ii) and their
integral approximants. For $E_0 = 0$ the sum over $m$ may be performed
exactly (see Section \ref{sec:e0=0}), and the McLaurin estimate has
been used only for the sum over $n$.

To understand the nature of the sum over $n$ it is useful to consider
the case $E_0 = 0$ and consider the contribution ${\tilde{\sigma}}_L(n)$ of a
given twist sector to the cross-section which is given by
(\ref{sigma-n-e0-0}). In Figure \ref{fig:averaging} we have plotted
$\sigma (n)$ versus $n$, both as histograms for integer $n$ and their
continuum approximations where $n$ is replaced by a real number.
\begin{enumerate}

\item{} The upper histogram (and the upper curve which is the 
continuum approximation) correspond to low resolution (high $R\
\Delta E$). In this case the $H(A\ n)$ term is negligible
and this yields the classical limit
\eq{sigma-class-n-e0-0} corresponding to the naive geometry
\eq{three}\footnote{Although we have used the typical value \eq{five}
in plotting the upper curve/histogram,
we could have used any other $N_n$ to arrive at the naive geometry.}.

\item{} The lower histogram corresponds to
high resolution (low $R\ \Delta E$) in which the $H(A\ n)$ term is
appreciable. As we have argued in Section \ref{fourfour}, this
histogram, for typical $N_n$, can be replaced by its continuum
approximation $[8 x/\sinh(\beta\ x)] \left(1+ H(A\ x)\right)$. Since
a {\em typical} $N_n$ is used, the lower curve represents 
{\em averaging} over the canonical (equivalently, microcanonical)
ensemble. 

\begin{figure}[ht!]
\centering
\includegraphics[scale=1]{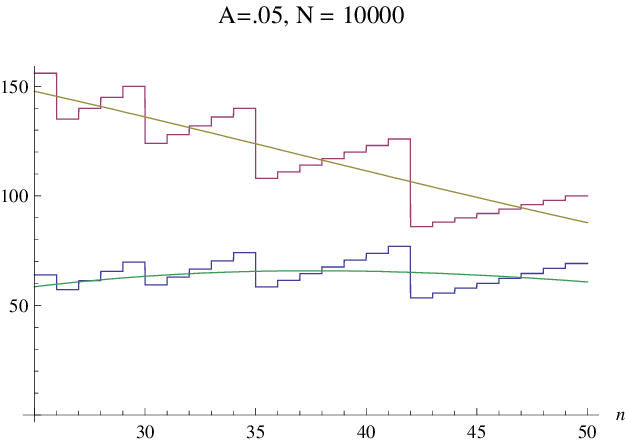}
\caption{We have plotted on the $y$-axis ${\tilde{\sigma}}_L(n)$ for
$E_0=0$, in units of $\pi\ R/(32 \ \Delta E)$.  The upper histogram and
curve refer to ${\tilde{\sigma}}_{L,{\rm classical}}(n)$ and its continuum
approximation, respectively. In other words the upper histogram
corresponds to $n\ [N_n]$ where $[\ ]$ represents nearest integer,
while the upper curve plots $8 x/\sinh(\beta\ x)$. The lower
histogram refer to the full expression, $n\ N_n \left(1+ H(A\
n)\right)$, while the lower curve plots the function $[8 x/\sinh(\beta\
x)] \left(1+ H(A\ x)\right)$ The upper curve corresponds to the naive
geometry which arises due to low resolution (large $A\equiv R\ \Delta E$) 
while the lower curve
refers to a different continuum which arises due to averaging over
microstates.}
\label{fig:averaging}
\end{figure}

The second continuum limit, obtained by averaging, yields
expressions for $\Delta\sigma_L$ of the same form as
we obtain in derivative corrected supergravity. 

\end{enumerate}  

\subsection{The Main results}

Let us summarize our main results of our computation of corrections
explained above:

\begin{enumerate}

\item{} 
The physically interesting regime (see Eq. \eq{regime}) is $1
\gg R\Delta E \gg 1/{\sqrt{N}}$. In this regime, the discreteness of
the system is manifest but for a {\em typical} state there are a large
enough number of energy levels in $\Delta E$ to give rise to a time
independent absorption cross-section.  
\subitem{(i)} 
When the central
value of the energy $E_0$ of the incident wave is much smaller than
the width $\Delta E$ we showed that the microstate dependent
correction due to discreteness has an upper bound which is {\em
independent of all energy scales} and proportional to $A_H\,\log(N)$
where $A_H$ is the area of the stretched horizon. Compared to the
microstate independent classical answer, this correction is suppressed
by a power of $1/(R\Delta E {\sqrt{N}})$.  

\subitem{(ii)} 
A better
bound has been obtained for $E_0 = 0$, the correction is bounded by
simply $A_H$ without the $\log (N)$ factor.  The sign of the
correction is {\em negative}. Note, however, that the total
cross-section is explicitly positive since it is a sum of positive
terms. Of course, in our approximation there is never a domain where
the correction is bigger than the classical term.
  
\subitem{(iii)} 
For general small
$E_0R$, we have  numerically 
calculated the difference between the cross-section
and its classical limit  for a typical state.  The results
show that this difference is {\em negative} and proportional to $A_H$.
\subitem{(iv)} 
For larger values of $E_0 R$ a numerical calculation
for the correction yields a {\em positive} term proportional to
$A_H$. However this is not in our regime of interest.

\item{} We have also obtained estimates for the correction 
for some {\em atypical} states. For the untwisted sector we found that
the bound on the finite $R$ correction {\em has the same $N$
  dependence as the classical result}. The correction is now
suppressed by a power of $1/(R\Delta E)$. For the maximally twisted
sector, the bound on the correction is independent of $N$. However
numerical results of the correction indicate that this is actually
suppressed exponentially in $N$ compared to the classical result.

\item{} The bounds for the correction for a Gaussian profile are
  similar to those for a Lorentzian profile. In particular, for
  typical states the bound is proportional to ${\sqrt{N}}$. The
  proportionality constant becomes a pure number for $E_0 \ll \Delta
  E$ and linear in $\frac{E_0}{\Delta E}$ for $E_0 \gg \Delta E$.
It is likely that there is a large class of energy profiles for which
  similar results will hold.

\end{enumerate}

\subsection{Appearance of Area of Stretched Horizon}

Perhaps the most significant aspect of our results is the fact that
{\em for typical states the correction to the naive geometry answer is
proportional to the area of the stretched horizon}.  This is because
the results of \cite{Dabholkar:2004yr,Castro:2007sd} indicate that
higher derivative terms lead to a modification of the geometry near
the singular horizon of the naive geometry. The modified geometry has
in fact a regular horizon whose area is exactly equal to the area of
the stretched horizon of the naive geometry, $A_H$.

It is tempting to speculate that the correction term somehow captures
this modified geometry with a horizon. We do not have a good
understanding of the bulk interpretation of such a correction. The
fact that the sign of the correction is negative in most cases we
studied numerically, also needs an understanding from the bulk
viewpoint, since the general result of \cite{Das:1996we} would seem to
suggest that the zero-frequency limit of the crosssection should be
given by the area of the geometry in the higher derivative theory and
should hence be a positive quantity
\footnote{The result may not apply,
though, if the scalar we considered ceases to remain minimally coupled
to gravity as we turn on the higher order corrections. For other black
holes, it has been found \cite{Moura:2006pz}
that higher derivative effects do change the
equation of motion for scalar modes. However, the cross-section still
remains proportional to the area with a nontrivial proportionality
constant involving the strength of the higher derivative term.
}.  
One
possible interpretation of our result\footnote{This possibility was
suggested to us by Samir Mathur.} is that the higher derivative
correction to the geometry is equivalent to putting a translucent wall deep
down the $AdS_3$ throat of the naive geometry. Since such a wall will
reflect part of the wave which has entered the $AdS_3$ throat, the
fraction which would enter this horizon will be less than the fraction
of incoming waves at infinity which entered the throat. This would
naturally explain why the corrected cross-section is smaller than the
classical cross-section, which measures the probability to enter the
throat.  If this is the case, it is not unlikely that the effect will
bring in a factor proportional to the area of the horizon of the
effective geometry, assuming such a geometry does indeed arise in our
situation due to higher derivative corrections to gravity. 
This is of course a vague explanation at this
point. We hope to form a more concrete picture in the future.

The space-time interpretation of the subleading corrections for {\em
atypical} states is, however, quite different. A given atypical state
corresponds to a specific microstate geometry and the microscopic
cross-section corresponds to the probability of entering the throat
region. Consider for example the microstates of the type analyzed in
sections (\ref{atypical1}) and (\ref{atypical2}).  The bulk
descriptions of these microstates have been discussed in Section 3.
As shown there, the leading answer for this quantity (in the regime
described by (\ref{lmathurregime1}) simply reproduces the answer in
the naive geometry. The calculations in \cite{Lunin:2001dt} and
\cite{Giusto:2004ip} give an answer valid in a less restrictive regime
where the second equation in (\ref{lmathurregime1}) is not
necessary. However since we are interested in finite $R$ corrections,
our calculation should be compared to a more refined calculation of
scattering of an incoherent beam with a finite energy resolution in
such geometries. We defer this calculation for future work.

\subsection{\label{sec:averaging}Averaging and Horizons}

A key feature of our results which is worth emphasizing is that for
sufficiently large $R(\Delta E)$, i.e. low energy resolutions, the
scattering experiment discussed in this paper perceives the naive
geometry for {\em any} microstate of the system \footnote{This is the
analog of the observation of \cite{balasub2} that correlation
functions in any microstate reproduce the AdS correlators for
sufficiently short times.}.  The effect of statistical coarse graining
over microstates is present in the subleading term. This term
certainly does not correspond to scattering in the naive geometry. If
anything, this corresponds to a geometry corrected by higher
derivative effects. If results similar to \cite{Dabholkar:2004yr,Castro:2007sd} hold in this case, the latter geometry has a regular horizon whose
Wald entropy agrees with the microscopic entropy.

\subsubsection{Comparison with the 3-charge D1-D5-P system}

This situation should be contrasted with leading order 
scattering from a D1-D5-P system with macroscopic amount of momentum in
the D1 direction as in
\cite{Callan:1996dv,Dhar:1996vu,Das:1996wn}. Here the relevant microstates
correspond to momentum states of the long string all moving in the
same direction (which we will call right moving). Now the probability
for absorption is a modification of
(\ref{transprob}) and (\ref{transprob2})
\ben {\cal
P}(E,T) = \sum_{n=1}^N 
N_n\,\frac{4 \kappa^2}{\pi\ V}\frac{1}{ER} \sum_{m=1}^\infty
(\frac{m}{nR})^2 \, \rho_P (m/nR)\,\left[ \frac{\sin
(E-\frac{2m}{nR})T}{(E-\frac{2m}{nR})} \right]^2
\label{transprob3charge}
\een
where $\rho_P (m/nR)$ is the distribution function of rightmoving
momentum among the quanta. This must satisfy
\ben
\sum_{m=1}^{M} \frac{m}{nR} \rho_P(m/nR) = P_R
\label{rhonormal}
\een
where $P_R$ is the total right moving momentum.

In the leading approximation
where the discreteness can be ignored, and for long time scales the
cross-section then becomes
\ben
\sigma_{3charge}(E) = \frac{\kappa^2 E}{4V}\, \sum_{n=1}^N \rho_P(E/2) 
nN_n
\label{crosscont3charge}
\een
The microstates are now specified by two distribution functions, $N_n$
for the twists and $\rho (p)$ for the momenta. 
The calculations of 
\cite{Callan:1996dv,Dhar:1996vu,Das:1996wn} were performed for a
typical state from the point of view of the ensemble defined by $\rho_P
(p)$. Then $\rho_P (p)$ has a thermal form
\ben
\rho_P (p) = \frac{1}{e^{p/T_R}-1}
\een
where $T_R$ is the right moving temperature which is related to
the total momentum $P_R$ by requiring (\ref{rhonormal}). Then for $E
\ll T_R$ one may approximate
\ben
\rho_P (E/2) \sim \frac{2T_R}{E}
\een
The factor of $E$ in the denominator then cancels the factor of $E$ in
the numerator in (\ref{crosscont3charge}), and using the fact $T_R$ is
proportional to the entropy of the system, which in turn is
proportional to the horizon area, one gets the result that the
cross-section is in fact exactly equal to the horizon area. 

Clearly, this result follows {\em only in a typical state}.  In other
words, {\em a statistical averaging over microstates has been
performed in addition to coarse-graining due to low resolution of
energies} to obtain this leading order result.

The key difference between the 2-charge system and the 3-charge
system in five dimensions is that the naive geometry of the latter has
a smooth large horizon (for large charges). This suggests a connection
between statistical averaging and horizons in the type of scattering
experiments we have considered in this paper.

\subsubsection{\label{BPS}
Comparison with scattering from massive heterotic BPS states}

In \cite{Mandal:1995qb,David:1996nm} scattering of a massless probe,
e.g. a graviton, from a massive heterotic BPS state
\cite{Sen:1995in} was considered.
The ``microstate'' of the BPS state is given by a polarization tensor
$\zeta$. There are many choices of $\zeta$ (called $\zeta_L$ in these
references), typically of a large rank, which all correspond to a
given mass $m$ and charges $\vec Q$.  The state of the probe is specified
by a polarization $\eta$ (of rank 1 or 2). Let us consider
for simplicity a process in which the polarizations $\zeta$ and
$\eta$ do not change. The structure of the 
elastic scattering amplitude (at small $\alpha't$) is given by
terms such as 
\bea
A(s,t,u) && \sim \frac{a}{\alpha' t} + b\ \eta_p\zeta_{pij..kl}
\zeta_{rij..kl}\eta_r  + O(\alpha' t) 
\nn
&& = \frac{1}{\alpha' t} \left( a + b\ \eta_p\zeta_{pij..kl}
\zeta_{rij..kl}\eta_r\ \alpha' t + o(\alpha' t)^2 \right)
\eea
where $a, b$ are numerical factors.
In \cite{Mandal:1995qb} it was shown that the leading term
in the small $t$ expansion matched exactly the Rutherford scattering
limit of the same probe scattered by the corresponding heterotic black hole 
\cite{Sen:1995in}. The subleading term, however, has a coupling
between the polarization states of the probe and the target BPS state.
It was noted in \cite{Mandal:1995qb} that

\begin{enumerate}

\item 
the low energy term $\sim 1/(\alpha' t)$ is independent of the
microstate and is reproduced by the ``naive'' heterotic black hole geometry,
and   
\item 
the subleading term depends on the ``microstate'' and violates
the no-hair property. 
\end{enumerate}

However, the new point we want to make here is that if we ``average''
over various ``microstates'', denoted in the sum below by
a label $q$ stuck to the polarization tensors $\zeta$, the
subleading term becomes
\[
\sum_q \eta_p\zeta^{(q)}_{pij..kl}
\zeta^{(q)}_{rij..kl}\eta_r = \eta_p \eta_r \delta_{pr} = 1
\]
leading to a new `geometric' (microstate-independent) expression
\[
A(s,t,u) \sim \frac{1}{\alpha' t} 
\left( a + b\alpha' t + o(\alpha' t)^2 \right)
\]
It may be interesting to note the similarity to the
scattering off the 2-charge D1-D5 system  we consider in this paper.
It is interesting to find out if the microstate-dependent
subleading term can be understood in terms of some microstate
geometry. It will also be interesting to find out if the
geometric expression for the subleading term above
corresponds to a finite area horizon which arises out
of higher derivative corrections in the supergravity effective
action.

\section{\label{concl}Concluding Remarks}

We have performed a detailed analysis of the microstate dependence of
scattering from the D1-D5 system in five dimensions where the internal
space is $T^4$. The analysis can be extended to the case where the
internal manifold is a $K3$.
Moreover, microscopic
calculations of the absorption cross-sections for a large class of
string theory black holes are almost identical to the present
case. This suggests that one should perform a similar study for other
black holes. In particular the 3-charge system in five dimensions is a
straightforward generalization of the 2-charge system (the main
formula is given above in (\ref{transprob3charge}). Our methods can be
easily adapted to this case. It would be interesting to see whether
corrections to the classical answer (which again comes from the
difference of the expression (\ref{transprob3charge}) and its integral
approximation) reflect the change of the geometry due to higher
derivative supergravity effects.  Our considerations can be also
directly applied to the 3- and 4- charge systems in four dimensions.

The outstanding open problem here which needs to be addressed in future
work is the bulk understanding of our microstate dependent
corrections. For typical states, we have offered a very qualitative
speculation in this regard, but clearly a lot more work is needed to
obtain a precise correspondence. It is also important to understand
the relationships of the corrections we have computed for specific
microstates to more refined calculations of wave propagation in
microstate geometries, as discussed in the previous section.

Finally we emphasize that all our results are in the orbifold limit of
the CFT. It is important to find out possible modifications of the
result in the presence of deformations.

\subsection*{Acknowledgements}

We have benefited from many discussions with Rajiv Bhalerao, Atish
Dabholkar, Avinash Dhar, Justin David, K.T. Joseph, Shiraz Minwalla,
Ashoke Sen and Spenta Wadia. We are indebted to Samir Mathur for
numerous discussions at all stages of this work and for generously
sharing his insight. We are grateful to Justin, Samir, Shiraz and
Spenta for their comments on the manuscript. We also thank Per Kraus
for a correspondence clarifying his work.  Much of this work was
performed during the Monsoon Workshop on String Theory held at
TIFR. We wish to thank International Center for Theoretical Sciences
(ICTS), TIFR and the organizers of this workshop for providing an
extremely stimulating environment. The work of S.R.D was partially
supported by a National Science Foundation Office of International
Science and Engineering grant as a part of NSF-PHY-0555444 and by
ICTS. Finally we would like to thank the people of India for
supporting research in fundamental physics.

\appendix

\section{Absorption cross-section in the naive geometry}
\label{classical1}

Consider the wave equation (\ref{s-wave})
\be
[(f_1 f_5)^{1/2}w^2 
+ \frac1{r^3 f_5}\del_r(r^3 f_5(f_1 f_5)^{-1/2}\del_r] S(r)=0
\label{s-wave1}
\ee

\mysec{Far region}

The geometry of the far region is flat space-time, and
the equation becomes
\be
[w^2 + \frac1{r^3} \del_r{r^3 \del_r}] S=0
\ee
The solution is
\be
S(r)= \frac1{\rho}[ A J_1(\rho) + B Y_1(\rho)]
\ee
where $\rho=wr$. To avoid complications coming from integer order
Bessel functions we will consider in fact the general case where the
transverse dimension is $(q+1)$. Then the solution is given by
\ben
S(r) = \frac{1}{\rho^\nu} \left[ A\,J_\nu (\rho) + B\,J_{-\nu} (\rho)
  \right] 
\label{farsol}
\een
where $\nu = \frac{1}{2}(q-1)$ is assumed to be non-integral.
At $\rho \to \infty$, using the standard asymptotics for Bessel
functions we get
\ben
S(r)
\sim \frac{1}{\rho^{\nu + \frac{1}{2}}} \left[ C\, e^{i\rho} + D\,
  e^{-i\rho} \right]
\een
where 
\ben
\frac{C}{D} = e^{-i\frac{\pi}{2}}\frac{1 + 
\frac{B}{A}e^{i\pi\nu}}{1 + \frac{B}{A}e^{-i\pi\nu}}
\een
The probability of absorption is then given by
\be
{\cal P}= 1 - |\frac CD|^2 = \frac{4 {\rm
    Im}\,(\frac{B}{A})\,\sin(\pi\nu)}{1+ |\frac AB|^2 +
  \frac{B}{A}e^{-i\pi\nu} + \frac{B^\star}{A^\star}e^{-i\pi\nu}}
\label{prob1}
\ee

\mysec{Near region}

In the near region, defined in \eq{regions}, we get the (Poincare
patch of the) $AdS_3 \times S^3 \times T^4$ geometry, and \eq{s-wave}
becomes
\be
[w^2 + \frac{r}{\ell^4} \del_r{r^3 \del_r}]S=0
\ee
As above, we will consider the generalized case of $AdS_{p+2} \times S^q
\times T^4$ so that the equation becomes
\be
[w^2 + (\frac{\ell}{r})^{p-2} \del_r{(\frac{r}{\ell})^{p+2} \del_r}]S=0
\ee
The surface $r=0$ is the ``Poincare horizon''. We will
consider solutions which are ingoing at $r=0$. This is given
by
\be
S(r)= E\,( \frac{w \ell^2}{r} )^\mu\,H^{(1)}_\mu (\frac{w
  \ell^2}{r})~~~~~~~~~~\mu = \frac{p+1}{2}
\label{near}
\ee
In the limit $r \rightarrow 0$ this solution behaves as
\ben
S \rightarrow e^{i\frac{w \ell^2}{r}}
\een

\mysec{Intermediate Region}

In the intermediate region, we have to use the large $r$ behavior of
(\ref{near}) and the small $r$ behavior of (\ref{farsol}) and match
the two solutions. The small $r$ behavior of (\ref{farsol}) is
\ben
S \sim \frac{A}{2^\nu\,\Gamma(1+\nu)} +
\frac{B\,2^\nu}{\rho^{2\nu}\,\Gamma(1-\nu)} 
\label{farnear}
\een
while the large $r$ behavior of (\ref{near}) is
\ben
S \sim \frac{iE}{\sin(\pi\mu)} \left[ \frac{(w
    \ell^2)^{2\mu}\,e^{-i\pi\mu}}{r^{2\mu}\,2^\mu\,\Gamma (1+\mu)} -
  \frac{2^\mu}{\Gamma (1-\mu)} \right]
\label{nearfar}
\een

\mysec{Matching and Cross-section}

The expressions (\ref{nearfar}) and (\ref{farnear}) can be matched if
$\mu = \nu$, i.e. $q = p+2$. For our present case $p = 1$, so that
$\mu=\nu=1$. However, we will retain a general $\mu,\nu$. Matching
then yields
\ben
\frac{B}{A} = -
\frac{e^{-i\pi\mu}}{2^{4\mu}\,[\Gamma(1+\mu)]^2}\,
[\Gamma(1-\nu)]^2\,(w\ell)^{4\mu}  
\een
Using $\mu=\nu$ this give the following expression for the probability
of absorption (\ref{prob1})
\ben
{\cal P} =
\frac{4\pi^2}{[\Gamma(\mu)]^2}\frac{(w\ell)^
{4\mu}}{2^{4\mu}\,[\Gamma(1+\mu)]^2} 
\label{prob2}
\een
The cross-section $\sigma_{cl}$ is obtained by multiplying this quantity by the
fraction of a spherical wave which is in a plane wave. For a $q+1$
dimensional transverse space this gives 
\ben
\sigma_{cl} = {\cal P}\,(4\pi)^{\frac{q-1}{2}}\,\Gamma (\frac{q+1}{2})
\frac{1}{w^q}
\een
Substituting (\ref{prob2}) we finally get
\ben
\sigma_{cl} = \frac{4\pi^2 (4\pi)^\mu}{[\Gamma(\mu)\Gamma(1+\mu)]^2
  \frac{\Gamma(1+\mu)}{2^{4\mu}}w^{2\mu-1}\,\ell^{4\mu}} 
\een
For our present case, $\mu=\nu=1$ yields
\ben
\sigma_{cl} = \pi^3 \ell^4\,w
\een
which is (\ref{classcross}).

\section{\label{integral-approx} McLaurin integral approximation
for sums}

The McLaurin integral approximation states that 
if a function $f_L(x)$ is positive
and monotonically decreasing in $P \le x \le Q$,
the following is true
\be
\int_P^Q dx\ f_L(x) + f_L(P) >\sum_{n=P}^{Q}  f_L(i) > \int_P^Q dx\ f_L(x) + f_L(Q), 
\ee
We can rewrite the above in the form of an estimate for the
sum:
\be
\sum_{n=P}^{Q}  f_L(i) 
=\int_P^Q dx\ f_L(x) + f_L(Q) + \eta_1 \left( f_L(P)- f_L(Q) \right),\;
0<\eta_1< 1
\label{mclaurin-decreasing}
\ee
Similarly, for a positive monotonically increasing function $f_L(x)$
in $P' \le x \le Q'$, we get
\be
\sum_{n=P'}^{Q'}  f_L(i) = \int_{P'}^{Q'} dx\ f_L(x)
+ f_L(P') + \eta_2 \left( f_L(Q')- f_L(P') \right),\;
0<\eta_2< 1
\label{mclaurin-increasing}
\ee
Using these two results we can find integral estimates for
sums, for any function $f_L(x)$ with a finite number of minima and maxima,
as we will do now.

\subsection{\label{sec:sum-m}Details of 
estimation of $\Delta{\tilde{\sigma}}_L(n)$}

We will use  the notation 
\[I(P,Q)=\int_P^Q dx\, f_L(x),
S(P,Q)= \sum_{m=P}^Q f_L(m).\]

Let us apply the McLaurin integral estimates
above to \eq{del-sig-n} which we rewrite as
\be
\Delta {\tilde{\sigma}}_L(n) 
= \frac{(nR)^2}{16} [S(0,\infty)
- I(0,\infty)]
\label{del-sig-n-app}
\ee

In $ 0< x < \infty$ the function $f_L(x)$ is positive
and  has one extremum, {\it viz.} 
a maximum at
\be
x_1=\sqrt{a^2 + b^2}
\label{f-max-x}
\ee
Thus, $f_L(x)$ is monotonically increasing in 
the segment $x \in (0,x_1)$
and monotonically decreasing in the segment  
$x \in  (x_1, \infty)$. 

In segment $(0,n_1) \subset(0,x_1)$ 
we use \eq{mclaurin-increasing} and obtain
\be
 S(0,n_1) = I(0,n_1) + \eta_1\, f_L(n_1), \; 0< \eta_1< 1 
\label{mclaurin-increasing-1}
\ee
Here $n_1 = \lfloor x_1 \rfloor$ denotes the
integer part of  $x_1$.

In segment $(n_1+1, \infty) \subset
(x_1, \infty)$ we use \eq{mclaurin-decreasing} and 
have (using $f_L(\infty)\to 0$)
\be
S(n_1 + 1, \infty)= I(n_1+1, \infty)+ 
\eta_2\, f_L(n_1+1), \; 0< \eta_2< 1 
\label{mclaurin-decreasing-1}
\ee
Combining the last two equations, we get
\be
S(0,\infty)= I(0,\infty)  - I(n_1,n_1+1) + \eta_1\, f_L(n_1)
+  \eta_2\, f_L(n_1+1)
\ee
We can approximate\footnote{Subleading corrections to
this approximation are given by replacing 
$f_L(x)= f_L(x_1) + f''(x_1)(x-x_1)^2/2 + ...$ and are down by additional
factors of $O(1/a^2)$ for large $a$.} $I(n_1, n_1+1) 
\approx f_L(n_1)\approx f_L(n_1+1) \approx f_L(x_1)$.
Thus,
\be
S(0,\infty)- I(0,\infty) \approx f_L(x_1)\ \eta_3,
\;\; \eta_3= \eta_1 +\eta_2-1 \in (-1,1)
\label{error-m}
\ee
By using this in \eq{del-sig-n-app}, and the value
\be
f_L(x_1) = \frac{b^2 + a^2}{(a^2 + (\sqrt{b^2 + a^2} - b)^2)^2},
\label{mclaurin-m}
\ee
with $(a,b)$ as in \eq{def-f}, we obtain
\be
\Delta{\tilde{\sigma}}_L(n) 
\approx \eta_3(n) \frac1 4 
\frac{E_0^2 + {\Delta E}^2}{({\Delta E}^2 + (\sqrt{E_0^2 + {\Delta E}^2}
 - E_0)^2)^2}
\label{error-sigma}
\ee
We have 
displayed the possible dependence of the fraction $\eta_3$
on $a,b$ and hence on $n$.

\section{\label{gaussian} Results for a Gaussian Profile}

In this section we give the results for the upper bound on the
correction to the cross-section when we use a Gaussian profile of the
form given in (\ref{defprofiles}).

In this case, the absorption cross-section $\sigma_L
(E_0,\Delta E)$ becomes, instead of (\ref{kthree}),
\bea
\sigma_G (E_0, \Delta E)& =& 
\frac{2 \kappa^2}{ VR}\,K_G(E_0,\Delta E)\,
\sum_{n=1}^N N_n  {\tilde{\sigma}}_G(n), 
\nn
{\tilde{\sigma}}_G(n) &\equiv& \sum_{m=1}^\infty
(\frac{m}{nR})^2
{\rm exp} \left[ - \frac{(\frac{2m}{nR}-E_0)^2}{(\Delta E)^2} \right] 
\label{kthreeg}
\eea
where the function $K_G(x)$ is the normalization defined in
(\ref{defkrho}) for the Gaussian profile
\ben
K_G(E_0,\Delta E)  = \frac{2}{(\Delta E)^2}~ \tilde K_G(E_0/\Delta E)
\label{class3-1-g}
\een
and $\tilde K_G(x)$ has been defined in (\ref{def-k-l-gaussian}).
When the sum over $m$ is replaced by an integral we once again recover
the classical answer (\ref{lorentzclass1}), exactly as in 
equation (\ref{sumoverm})-(\ref{kfour}). The function which appears in
the integral is now $f_G(x)$ (instead of $f_L(x)$ of (\ref{def-f})),
\ben
f_G(x) = x^2~{\rm exp} \left[-\frac{(x-b)^2}{a^2} \right]
\label{def-f-g}
\een
where $a$ and $b$ are defined in (\ref{def-f}).

The departure from the classical limit is now given by
\bea
\Delta \sigma_G  (E_0, \Delta E)& =&
\sigma_G (E_0, \Delta E) - {\sigma}_{G,\rm classical} (E_0, \Delta E)
\nn
& = & \frac{2 \kappa^2}{ VR}\,K_G(E_0,\Delta E)\,
\sum_{n=1}^N N_n \Delta {\tilde{\sigma}}_G(n)
\label{del-sig-g}
\eea
where
\be
\Delta {\tilde{\sigma}}_G(n) \equiv {\tilde{\sigma}}_G(n) - 
{\tilde{\sigma}}_{G,{\rm classical}}(n)
= \frac{1}{(nR)^2}\left( \sum_{m=0}^\infty f_G(m) - \int_0^\infty\ dx\ f_G(x)
\right)
\label{del-sig-n-g}
\ee

The function $f_G(x)$ is quite similar to $f_L(x)$. In $0 < x <
\infty$ this is positive with a single maximum at $x=x_2$, 
\ben
x_2 = \frac{1}{2} \left[ b + {\sqrt{b^2 + 4a^2}} \right]
\een
The arguments of section (\ref{integral-approx}) then show that an
upper bound for $\Delta {\tilde{\sigma}}_G(n)$ is given in terms of
$f_G(x_2)$, exactly as in (\ref{error-m}). This leads to 
\ben
\Delta {\tilde{\sigma}}_G(n) \approx \frac{\eta_4(n)}{4} 
(\Delta E)^2 \left[ \left( \frac{E_0}{\Delta E} + {\sqrt{4 + 
\frac{E_0^2}{(\Delta E)^2}}} \right)^2~{\rm exp} \{
-\frac{1}{4} \left( {\sqrt{4 + 
\frac{E_0^2}{(\Delta E)^2}}} - \frac{E_0}{\Delta E} \right)^2 \}
\right]
\label{error-sigma-g}
\een
where $-1 < \eta_4 (n) < 1$.
Exactly as in (\ref{del-sigma-n1}), the factors of $(nR)$ have
cancelled and the only $n$ dependence is in $\eta_4 (n)$. 
Furthermore, the factor of $(\Delta E)^2$ in
(\ref{error-sigma-g}) cancels an overall factor of $1/(\Delta E)^2$
present in $K_G(E_0,\Delta E)$ in (\ref{class3-1-g})
leading to the final bound, which is the analog of
(\ref{sigma-bound}),
\ben
\Delta\sigma_{G,{\rm max}} (E_0, \Delta E) 
 =  
\frac{4\kappa^2}{VR}\, \tilde K_G(\frac{E_0}{\Delta E})\,
L_G(\frac{E_0}{\Delta E})\,
\sum_{n=1}^N N_n
\label{sigma-bound-g}
\een
where the function $L_G(x)$ is defined in (\ref{def-k-l-gaussian}).

As in the case of a Lorentzian profile, the microstate dependence of
this bound is entirely in the sum $\sum_{m=1}^N N_n$.
For a typical state this sum can be estimated as in
(\ref{sum-over-n}). This leads to the final expression for the bound
in (\ref{bounds-gaussian}).

For $E_0 \ll \Delta E$, we have the expansion
\ben
\tilde K_G(\frac{E_0}{\Delta E})\,
L_G(\frac{E_0}{\Delta E})
= \frac{4}{e} [1 + (2-\sqrt{\pi})\frac{E_0}{\Delta E} + \cdots]
\een
This means that for typical states we have
\ben
\Delta\sigma_{G,{\rm max}} (E_0, \Delta E) |_{E_0 \ll \Delta E} 
\propto A_H
\label{correctsmallE0-g}  
\een
just like the Lorentzian profile result (\ref{correctsmallE0-l}).
For $E_0 \gg \Delta E$
\ben
{\tilde{K}}_G (x) \sim \frac{1}{x}~~~~~~~~~L_G(x) \sim x^2
\een
so that 
\ben
\Delta\sigma_{G,{\rm max}} (E_0, \Delta E) |_{E_0 \gg \Delta E} 
\propto A_H~\frac{E_0}{\Delta E}
\label{correctlargeE0-g}  
\een
similar to (\ref{correctlargeE0-l}).

\end{document}